\documentclass[%
 reprint,
superscriptaddress,
 amsmath,amssymb,
  aps,
]{revtex4-1}

\usepackage[pdftex]{graphicx}
\usepackage{dcolumn}
\usepackage{bm}

\usepackage[version=3]{mhchem}
\usepackage{here}
\usepackage{ulem}

\begin{document}


\title{Nematicity in the pseudogap state of cuprate superconductors revealed by angle-resolved photoemission spectroscopy}

\author{S. Nakata}
\affiliation{Department of Physics, University of Tokyo, Bunkyo-ku, Tokyo 113-0033, Japan}
\author{M. Horio}
\affiliation{Department of Physics, University of Tokyo, Bunkyo-ku, Tokyo 113-0033, Japan}
\author{K. Koshiishi}
\affiliation{Department of Physics, University of Tokyo, Bunkyo-ku, Tokyo 113-0033, Japan}
\author{K. Hagiwara}
\affiliation{Department of Physics, University of Tokyo, Bunkyo-ku, Tokyo 113-0033, Japan}
\author{C. Lin}
\affiliation{Department of Physics, University of Tokyo, Bunkyo-ku, Tokyo 113-0033, Japan}
\author{M. Suzuki}
\affiliation{Department of Physics, University of Tokyo, Bunkyo-ku, Tokyo 113-0033, Japan}
\author{S. Ideta}
\affiliation{UVSOR Facility, Institute for Molecular Science, Okazaki 444-8585, Japan}
\author{K. Tanaka}
\affiliation{UVSOR Facility, Institute for Molecular Science, Okazaki 444-8585, Japan}
\author{D. Song}
\affiliation{National Institute of Advanced Science and Technology (AIST), Tsukuba 305-8568, Japan}
\author{Y. Yoshida}
\affiliation{National Institute of Advanced Science and Technology (AIST), Tsukuba 305-8568, Japan}
\author{H. Eisaki}
\affiliation{National Institute of Advanced Science and Technology (AIST), Tsukuba 305-8568, Japan}
\author{A. Fujimori}
\affiliation{Department of Physics, University of Tokyo, Bunkyo-ku, Tokyo 113-0033, Japan}
\affiliation{Department of Applied Physics, Waseda University, Shinjuku-ku, Tokyo 169-8555, Japan}

\date{\today}

\begin{abstract}
The nature of the pseudogap and its relationship with superconductivity are one of the central issues of cuprate superconductors.  Recently, a possible scenario has been proposed that the pseudogap state is a distinct phase characterized by spontaneous rotational symmetry breaking called ``nematicity''  based on transport and magnetic susceptibility measurements, where the symmetry breaking was observed below the pseudogap temperature $T^*$. Here, we report a temperature-dependent ARPES study of nematicity in slightly overdoped \ce{Bi_{1.7}Pb_{0.5}Sr_{1.9}CaCu2O_{8+$\delta$}} triggered by a uniaxial strain applied along one of the Cu-O bond directions. While the nematicity was enhanced in the pseudogap state as in the previous studies, it was suppressed in the superconducting state. These results indicate that the pseudogap state is characterized by spontaneous rotational symmetry breaking and that the nematicity may compete with superconductivity. These new experimental insights may provide clues for the nature of the pseudogap and its relation to the superconductivity.
\end{abstract}

\maketitle


\section{\label{sec:level1}Introduction}

The pseudogap in cuprate superconductors is characterized by the suppression of the density of states around the Fermi level below a characteristic temperature $T^*$ and above the superconducting transition temperature $T_\text{c}$.
Broadly speaking, two possible scenarios for the pseudogap have been discussed, that is, a precursor to the superconducting state and distinct order which competes with superconductivity \cite{Keimer2015From-quantum-ma}.
The latter scenario has been put forward by several measurements which detect distinct orders with spontaneous symmetry breaking such as translational and time-reversal symmetry breaking at or below the pseudogap temperature $T^*$ \cite{Ghiringhelli821,Chang2012Direct-obs,Comin1335,Le-Tacon2013Inelastic-,Comin390,PhysRevLett.96.197001,PhysRevB.89.094523,PhysRevLett.118.097003,0953-8984-26-50-505701,PhysRevLett.88.137005,Hinkov597,Lawler2010Intra-unit,Sato2017Thermodyna,Daou2010Broken-rot}. 

Recently, electronic nematicity that the electronic structure preserves the translational symmetry but breaks the rotational symmetry of the underlying crystal lattice has been found to exist in the pseudogap state \cite{PhysRevLett.88.137005,Hinkov597,Lawler2010Intra-unit,Sato2017Thermodyna,Daou2010Broken-rot,PhysRevB.92.224502,Wu2017Spontaneou}. 
As a possible mechanism, it has been proposed that the nematicity arises from fluctuations of stripe order \cite{Kivelson1998Electronic, RevModPhys.75.1201} or from the instability of the Fermi surface (so-called Pomeranchuk instability) \cite{doi:10.1143/JPSJ.69.2151, PhysRevB.73.214517, PhysRevB.64.195109, PhysRevB.72.024502,PhysRevLett.85.5162}. 
From experimental perspectives, the nematicity in the cuprate superconductors was first pointed out by transport measurements of lightly-doped \ce{La_{2-$x$}Sr_$x$CuO4 } and \ce{YBa2Cu3O_$y$} (YBCO) \cite{PhysRevLett.88.137005}. 
Anisotropic signals in the spin excitation measured by neutron scattering was observed for untwinned underdoped YBCO \cite{Hinkov597}. 
Inequivalent electronic states associated with the oxygen atoms in the $a$ and $b$ directions was detected by scanning tunneling spectroscopy measurements of underdoped \ce{Bi2Sr2CaCu2O_{8+$\delta$}} (Bi2212) \cite{Lawler2010Intra-unit}. {Nematic fluctuations in Bi2212 have been observed by Raman scattering \cite{Auvray2019Nematic-fl}.} Nernst effect and magnetic torque measurements on underdoped and optimally doped YBCO showed a systematic temperature dependence of the nematicity \cite{Daou2010Broken-rot,Sato2017Thermodyna}. 
The onset temperature of the in-plane anisotropic signals in the Nernst coefficient and that in the magnetic susceptibility coincide with $T^*$. {Furthermore the nematic susceptibility derived from elastoresistance experiments on Bi2212 diverges at $T^*$ \cite{doi:10.7566/JPSJ.89.064707}.
The order parameter-like behaviors of the nematicity in the Nernst and magnetic torque experiments \cite{Daou2010Broken-rot,Sato2017Thermodyna} and the divergence of the nematic susceptibility \cite{doi:10.7566/JPSJ.89.064707} indicate that the pseudogap state can be considered as a distinct thermodynamic phase characterized by the rotational symmetry breaking.}
Although the orthorhombic distortion of the \ce{CuO2} plane caused by the Cu-O chains along the $b$-axis of the untwinned YBCO samples already breaks the four-fold rotational symmetry even above $T^*$, the orthorhombicity is considered to help to form a microscopic nematic domain in one particular direction and enables us to detect nematicity in macroscopic measurements.
This weak orthorhombicity of the \ce{CuO2} plane for nematicity plays the same role as a weak external magnetic field for ferromagnets \cite{Vojta2009Lattice-sy}. 

Motivated by those previous studies on nematicity in the cuprates, we have performed angle-resolved photoemission spectroscopy (ARPES) measurements on slightly overdoped \ce{Bi_{1.7}Pb_{0.5}Sr_{1.9}CaCu2O_{8+$\delta$}} (Pb-Bi2212) ($T_\text{c} = 91$ K) by applying a uniaxial strain along the Cu-O bond direction to detect nematicity below $T^*$. At the doping level we chose, $T^*$ was not too high for ARPES experiments but {the divergence of the nematic susceptibility in Pb-Bi2212 was sufficiently strong \cite{doi:10.7566/JPSJ.89.064707}. }
While most of the studies on nematicity in the cuprate superconductors have been done on YBCO, Pb-Bi2212, whose \ce{CuO2} plane has the tetragonal ($C_4$) symmetry, is a more convenient material to study nematicity than YBCO.
Owing to the presence of a natural cleavage plane between the BiO layers and rich information accumulated from previous ARPES studies, Bi2212 is an ideal material for ARPES experiment to investigate novel phenomena in cuprates \cite{RevModPhys.75.473}. 
In the present work, we analyze the single particle spectra without assuming that the symmetry of the electronic structure is the same as the symmetry of the \ce{CuO2} plane to shed light on the possibly lowered symmetry of the electronic structure. 
Furthermore, using ARPES, one can investigate the nematicity not only in the normal and pseudogap states but also in the superconducting state, where transport and magnetic measurements cannot be performed due to the zero resistivity and the strong diamagnetism, respectively \cite{Sato2017Thermodyna,Daou2010Broken-rot,PhysRevB.92.224502}.


\section{\label{sec:level2}Experimental methods}

Pb-Bi2212 single crystals were grown by the floating-zone method. 
The hole concentration was slightly overdoped one ($T_\text{c} =$ 91 K) after annealing the samples in a \ce{N2} flow, which allowed us to measure samples above and below $T^*$ ($\sim$160 K) rather easily compared to the underdoped samples whose $T^*$'s are too high \cite{Vishik06112012}. 
We measured Pb-doped samples in order to suppress the superstructure modulation present in the BiO layers of Bi2212, which causes the so-called diffraction replicas of the Fermi surface shifted by multiples of $\vec{k} = \pm(0.21\pi ,0.21\pi)$ \cite{RevModPhys.75.473}. \par
YBCO has Cu-O chains in addition to the \ce{CuO2} plane, which helps nematic domains to align in one particular direction, while Pb-Bi2212 has no such an internal source of strain as the Cu-O chains and the in-plane crystal structure is tetragonal.  
Therefore, we applied a tensile strain to the sample along the Cu-O bond direction in attempt to align nematic domains in one direction using a device similar to that used for Fe-based superconductors, a picture and a schematic figure of which are shown in Supplementary Fig. S1 (c) \cite{PhysRevLett.115.027005}. Here, in analogy to ferromagnets, the tensile strain plays a role of a weak external magnetic field that align ferromagnetic domains along the field direction \cite{Vojta2009Lattice-sy}.
The strain was applied in air at room temperature, and the stressed sample was introduced into the ultrahigh vacuum and  cooled down to measurement temperatures before cleaving {\it{in situ}} (See Supplementary Information S1 for the strain estimated from the lattice parameters by x-ray diffraction measurement). Such an operation was necessary to obtain anisotropic signals from spectroscopic data because the size of the nematic domains is not necessarily larger than the beam size; otherwise, anisotropic signals may be averaged out \cite{Yi26042011,PhysRevB.90.121111}.
Note that in our experimental setup, it is impossible to detect nematicity which is diagonal to the crystallographic $a$ and $b$ directions \cite{Wu2017Spontaneou}, i.e., so-called diagonal nematicity \cite{Murayama2019Diagonal-n}\par

ARPES measurements were carried out at the undulator beamline BL5U of UVSOR using an MBS A-1 analyzer. The photon energy $h\nu$ was fixed at 60 eV. The energy resolution was set at 30 meV. The linear polarization of the incident light was chosen perpendicular to the analyzer slit and the tilt axis parallel to the analyzer slit, which realizes the unique experimental configuration that preserves the equivalence of the $a$- and $b$-axis directions with respect to the light polarization $\vec{E}$ and the strain direction, as shown in the inset of Fig. 1. This setting guarantees that the anisotropy of the spectroscopic data between the $a$- and $b$-axis directions does not originate from  matrix-element effects but from intrinsic inequivalence in the electronic structure. The measurements were performed in the normal, pseudogap, and superconducting states in two series, that is, with increasing temperature and with decreasing temperature in order to check the reproductivity. The samples were cleaved {\it in situ} under the pressure better than $2\times 10^{-8}$ Pa.

\begin{figure}[h]
\includegraphics[width=8cm]{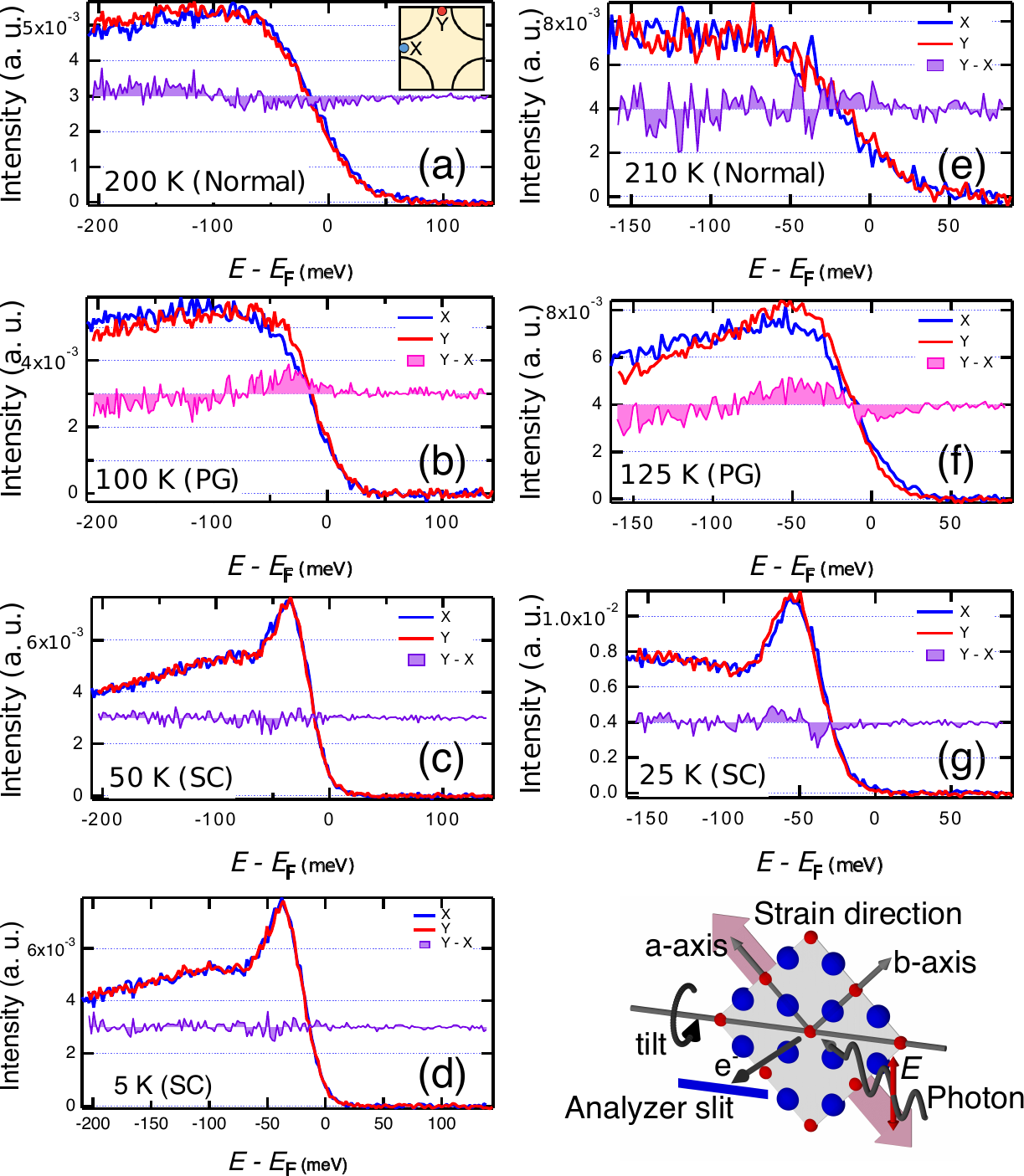}
\caption{Energy distribution curves (EDCs) of Pb-Bi2212 at various temperatures. The measurements were performed with decreasing temperature for (a)-(d) and increasing temperature for (e)-(g). At each temperature, EDCs at the X and Y points and the difference between them are displayed. Note that the X (Y) point is slightly off from $(-\pi, 0)$ ($(0, \pi)$) to prevent the overlap of diffraction replicas (see also supplementary information S6). To discuss the line shape of the EDCs at each temperature, {the spectral weight integrated in the displayed energy range has been normalized.} In the pseudogap state ($T =$ 100 K and 125 K), the spectral intensities between the X and Y points are clearly different while they are identical in the normal ($T =$ 200 K and 210 K) and superconducting states ($T =$ 5 K, 25 K, and 50 K). Inset shows the experimental geometry including the crystallographic axes, the strain direction, light polarization, and the analyzer slit, which ensure the spectroscopic equivalence of the $a$- and $b$-axis directions.}
\label{fig:EDC}
\end{figure}

\begin{figure*}[t]
\includegraphics[width=17cm]{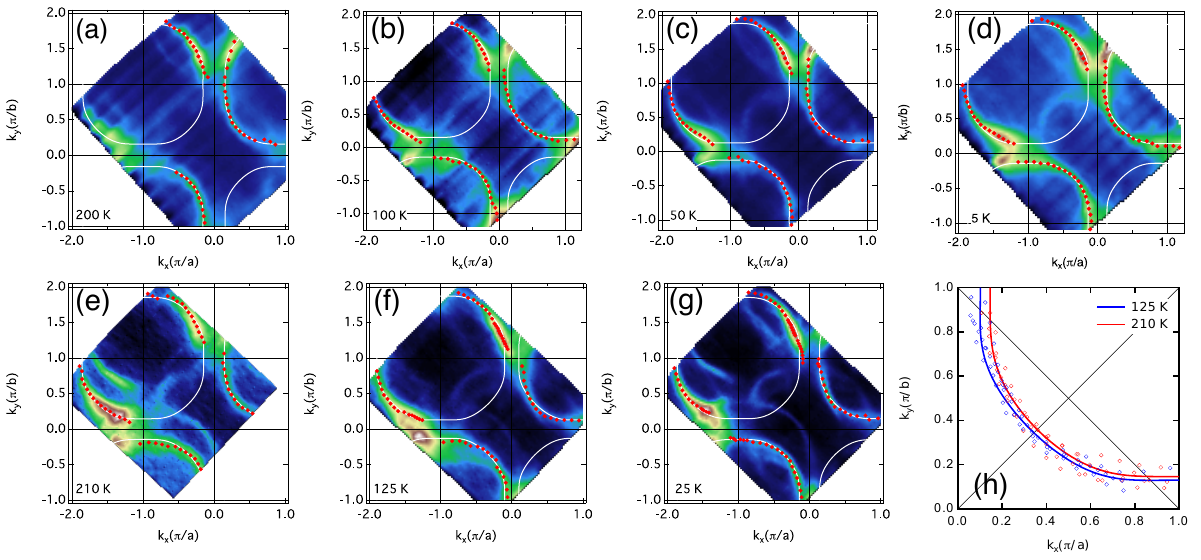}
\caption{Constant-energy-surface mapping of the ARPES spectra of Pb-Bi2212  taken at various temperatures with decreasing temperature (a)-(d) and increasing temperature (e)-(g). The constant energy is $E-E_\text{F}$ = - 30 meV and the intensity is integrated within $\pm$ 10 meV of $E-E_\text{F}$ = - 30 meV. Red circles show the constant-energy surface determined from the peaks of the momentum distribution curves (MDCs). White curves are the fitted curves using the tight-binding model of Eq. (2). In (e)-(g),  extrinsic Fermi-surface features which most likely arose from another tilted cleaved surface can be seen, but have been separated out (as described in detail in Supplemental Information S6) and ignored in the present analyses. {(h)  Constant-energy surface at 125 K and 210 K folded in the first quadrant of the first Brillouin zone and fitted curves in a range allowed by the symmetry.}}
\label{fig:Mapping}
\end{figure*}

\section{\label{sec:level3}Results and Disucussion}

In Fig. 1, energy distribution curves (EDCs) around $\vec{k} = (-\pi,0)$ (X point) and $(0,\pi)$ (Y point) at various temperatures are shown. The data were obtained from one sample but from two cleavages: on one cleaved surface, we performed ARPES measurements with cooling the sample at $T =$ 5 K, 50 K, 100 K, and 200 K and, on the other cleaved surface, with heating the sample at $T =$ 25 K, 125 K, and 210 K. In both measurements, the tensile strain was applied in the $x$-direction ($||a$-axis) as shown in the inset of Fig. 1. In the cooling series, $T$ = 200 K $\rightarrow$ 100 K $\rightarrow$ 50 K $\rightarrow$ 5 K, in the normal state (200 K), the line shapes of the EDCs were almost identical between the X and Y points, as  expected from the four-fold rotational symmetry of the \ce{CuO2} plane. With decreasing temperature, the line shapes of the EDCs around the X and Y points became different below $T^*$ in the pseudogap state (100 K) and then identical again below $T_\text{c}$ in the superconducting state (5 K and 50 K).  In the heating series, $T$ = 25 K $\rightarrow$ 125 K $\rightarrow$ 210 K, the spectral changes of the cooling series was reproduced as shown in Fig. 1.
Let us focus on the pseudogap state, where difference in the line shapes of the EDCs is present between the X and Y points. In Bi2212, it is well known that the Cu-O band is split into the anti-bonding and bonding bands due to the bilayer structure \cite{PhysRevLett.86.5550}. {From the dispersions near $\vec{k} = (\pi,0)$ in  overdoped sample ($T_\text{c}=$ 91 K), whose doping level is the same as the present sample, we consider that the energies of the bottoms of the anti-bonding band and bonding band are located around $E-E_\text{F} =$ -25 meV and -110 meV, respectively, in the pseudogap state. (See Supplementary Information S2.)} Thus, we conclude that the intensity of the bonding (anti-bonding) band is higher around the X (Y) point than that around the Y (X) point, reflecting the in-plane anisotropy of the electronic structure in the pseudogap state under the uniaxial strain. Here, we would like to emphasize that this inequivalence between the X and Y points is not caused by matrix-element effect because the $x$- and $y$-directions are equivalent for a tetragonal sample in the present measurement geometry {(See also Supplementary Information S3)}.

In Fig. 2 (a)-(g), the intensity maps of the constant-energy surface at $E-E_\text{F} =$ -30 meV at various temperatures are displayed. {We have chosen $E-E_\text{F} =$ -30 meV rather than $E_\text{F}$ in the following analysis. This is because the dispersion of quasiparticles near $E_\text{F}$ is partially gapped due to the pseudogap and superconducting gap opening at low temperatures.} The intensity around $k_y=-k_x$ almost vanishes due to matrix-element effect (See Supplementary Information S4 for the matrix-element effect). Note that additional features which resemble the constant-energy surface but are shifted in $k$ space can be seen in the data of the heating series (Fig. 2 (e)-(g)). They are due to extrinsic signals arising from a tilted surface that appeared when the sample was cleaved at low temperature. Well-known diffraction replica shifted by $(\pi,\pi)$ was also observed. (See Supplementary Information S5 and S6) However, the extrinsic features could be clearly separated from the intrinsic ones in the momentum distribution curves (MDCs), and thus the extrinsic features do not affect the present analysis. At  $\vec{k} = (\pi,0)$ for $h\nu =$ 60 eV, the intensity of the anti-bonding band is only a little weaker than the bonding band but not negligible \cite{PhysRevLett.89.077003,PhysRevLett.90.207001}, and their MDC widths are large ($\sim0.3 \ \pi/a$ FWHM) compared to the momentum separation of the two bands ($\sim0.1 \ \pi/a$). Therefore, we extracted the constant-energy surface practically as a single band rather than multiple bands, i.e., overlapping anti-bonding and bonding bands. The constant-energy surface has been determined from the peak positions of the MDCs fitted using Lorentzians.
In Fig. 2 (h), the constant-energy surface at 125 K and 210 K folded in the first quadrant of the first Brillouin zone are overlaid on each other and compared. The constant-energy surface around the Y point at 125 K is closer to $(0,\pi)$ than that at 210 K. This is qualitatively consistent with the fact that the intensity of the anti-bonding (bonding) band becomes strong near the Y (X) point in the pseudogap state as displayed in Fig. 1 (f). Therefore, we interpret the intensity transfer from the bonding band to anti-bonding band as one goes from the X point to the Y point shown in Fig. 1 (f) makes the constant-energy surface distorted around the Y point. 

To understand the anistropy of the constant-energy surface more quantitatively, we have estimated it with the tight-binding model in the following way.
The band dispersion near the Fermi level of the cuprates can be fitted using the standard tight-binding model
\begin{eqnarray}
	\varepsilon (k_x,k_y)- \mu = \varepsilon_0 &-&2t(\cos k_x + \cos k_y) \nonumber\\
	-4 t' \cos k_x \cos k_y &-&2t'' (\cos 2k_x + \cos 2k_y)  , \label{eq:3.2.1}
\end{eqnarray}
where $t$, $t'$, and $t''$ are the nearest-neighbor, second-nearest-neighbor, and third-nearest-neighbor hopping parameters, respectively, and $\mu$ is the chemical potential \cite{PhysRevB.95.075109}. The model has four-fold rotational ($C_4$) symmetry. 
In order to examine the possibility of the $C_4$ rotational symmetry breaking into $C_2$ symmetry, we introduce an anisotropy parameter $\delta$ which represents the orthrombicity of the hopping parameters ($t$ and $t''$) \cite{PhysRevB.73.214517} as
\begin{eqnarray}
&&\varepsilon (k_x,k_y) - \mu=\varepsilon_0 -2t[(1-\delta)\cos k_x +(1+\delta) \cos k_y] \nonumber\\
&& -4 t' \cos k_x \cos k_y -2t'' [(1-\delta)\cos 2k_x +(1+\delta) \cos 2k_y] .\nonumber\\ \label{eq:3.2.2}
\end{eqnarray}
By fitting the constant-energy surface using Eq. (\ref{eq:3.2.2}), we have estimated its deviation from the $C_4$ symmetry through the finite $\delta$ values {(For detailed fitting procedures, see Supplementary Information S7)}.

The temperature evolution of the anisotropy parameter $\delta$ thus derived is shown in Fig. 3. In the normal state ($T > T^*$), $\delta$ is close to zero, in both heating and cooling series, which was expected from the four-fold rotational symmetry of the \ce{CuO2} plane. It is also consistent with the EDCs in Figs. 1 (a) and (e). In the pseudogap state ($T_\text{c} < T < T^*$), however, $\delta$ became finite. The sign of $\delta$ indicates that the hopping parameters along the tensile strain direction became small in the pseudogap state. The sign of $\delta$ is also the same as the anisotropic Fermi surface of \ce{YBa2Cu4O8} under the uniaxial strain from the Cu-O chains leading to the small difference between the $a$ and $b$ lattice constants \cite{PhysRevB.80.100505}. Thus, the uniaxial strain seems to serve as an external perturbation to align a majority of nematic domains in one direction (See also Supplementary Information S8 for the effect of strain on the magnitude of the hopping parameters). Figure 3 also shows that $\delta$ is suppressed in the superconducting state. 
In the normal and pseudogap states, our result is consistent with the previous magnetic and transport measurements on YBCO in that the nematicity becomes finite below $T^*$\cite{Sato2017Thermodyna,Daou2010Broken-rot}. More concretely, for example, the anisotropy of the magnetic susceptibility of YBCO, which was derived from the magnetic torque measurements \cite{Sato2017Thermodyna} is as large as 0.5$\%$ just above $T_\text{c}$. Although further studies are necessary to compare the magnetic susceptibility and the single-particle spectral function quantitatively, we naively believe that the $\delta$ whose order of the magnitude was 1\% in our measurements would have high impacts on various physical properties such as charge and/or magnetic susceptibilities.
As for the superconducting state, the magnetic and transport measurements cannot give any information, because of the giant diamagnetism and zero resistivity, respectively, about the underlying electronic structures while our ARPES result indicates possible competition between nematicity and superconductivity.


From the theoretical side, dynamical mean-field theory (DMFT) combined with the fluctuation exchange (FLEX) approximation for the Hubbard model has shown that a Pomeranchuk instability, where the $C_2$ anisotropy of the Fermi surface is induced, appears in the overdoped region but coexists with superconductivity \cite{PhysRevB.95.075109}. In contrast, according to a cellular dynamical mean-field theory (CDMFT) study for the Hubbard model, the $C_4$ symmetry breaking was shown in the underdoped pseudogap regime rather than in the overdoped regime \cite{PhysRevB.86.094522,PhysRevB.82.180511}. 
According to mean-field calculations of the $t-J$ model, the Pomerachuk instability competes with superconductivity \cite{doi:10.1143/JPSJ.69.2151,PhysRevB.75.155117}. Thus, it is not clear at present whether the Pomeranchuk instability consistently explains our result or not. 

Nematicity has also been considered as arising from fluctuations of charge-density wave (CDW) or stripes \cite{Nie03062014,PhysRevB.92.174505}. In YBCO, CDW was observed by x-ray scattering experiments and found to reside inside the pseudogap phase and competes with superconductivity \cite{Ghiringhelli821,Chang2012Direct-obs,Comin1335,Le-Tacon2013Inelastic-}. CDW was also found in Bi2212 and to compete with superconductivity \cite{PhysRevB.89.220511,Silva-Neto2014Ubiquitous} as in the case of YBCO. More recent work on underdoped Bi2212 ($T_\text{c} =$ 40 K) showed that the elastic peak that represents the CDW survived up to  $T^*$ \cite{Chaix2017Dispersive}. From the fact that static CDW in Bi2212 competes with superconductivity and possibly survives up to $T^*$ including our case, it is also likely that the nematicity observed in our measurements arises from fluctuating CDW.
Furthermore, there have been recent theoretical attempts to explain the nematicity using by the pair-density-wave (PDW) model \cite{PhysRevB.97.174511,Tu2019Evolution-}. As shown in the calculated single-particle spectra in \cite{Tu2019Evolution-}, the inequivalence between the X and Y points may be attributed of the intensity difference predicted for incommensurate PDW states at finite temperatures.\par
{In Fe-based superconductors, nematicity has been observed as the different populations of different orbitals (such as $yz \ vs \ zx$ orbitals) revealed by ARPES experiments \cite{Yi26042011}. In the present study, the observed anisotropy might be due to the different populations of the oxygen $p_x$ and $p_y$ orbitals because the entire Fermi surface consists of the copper $3d_{x^2-y^2}$ orbital hybridized with the oxygen $p_x$ and $p_y$ orbitals \cite{PhysRevLett.114.257001}. However, nematicity can also be triggered by other mechanisms, e.g., the Pomeranchuk instability in single band systems and, therefore, it is not clear whether the anisotropy of the orbital populations is important to explain the nematicity observed in the cuprates as in the case of the Fe-based superconductors. Although at present it is difficult to identify the microscopic origin of the nematicity, our result provides evidence that the pseudogap state shows nematicity and competes with superconductivity. In contrast, the STM experiments have indicated nematicity even in the superconducting state \cite{Fujita612}. In order to reconcile the STM experiments with the apparent competition between the nematicity and the superconductivity implied in the present study, there is the possibility that the nematicity is masked by the $d$-wave superconducting order parameter, whose amplitude should be identical between the two Cu-O bond directions, in the present ARPES data.}

\begin{figure}[t]
\includegraphics[width = \textwidth/2-1cm]{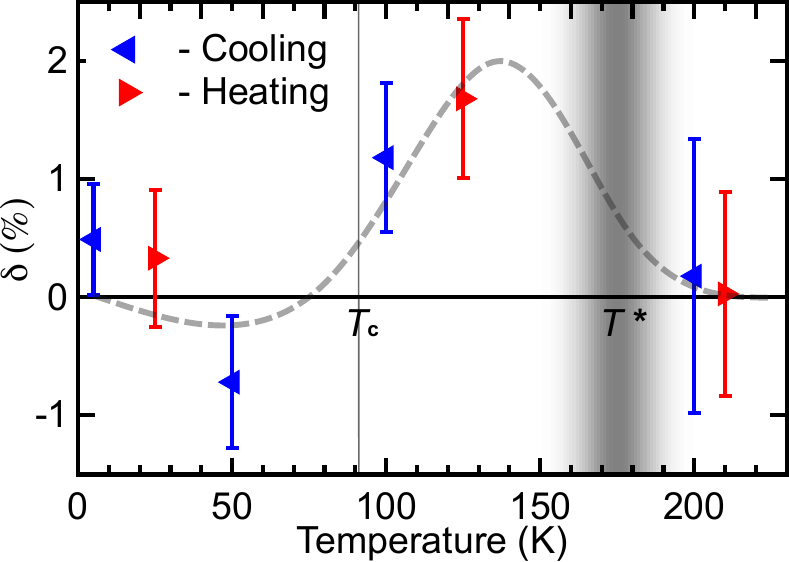}
\caption{\label{fig:delta} Temperature dependence of nematicity represented by the anisotropy parameter $\delta$ defined by Eq. (2). The error bars are standard errors of the fitting parameter at each temperature. The dashed gray line is a guide to the eye.}
\end{figure}


\section{\label{sec:level4}Conclusions}

{In conclusion, we have demonstrated the presence of nematicity in the pseudogap state of Pb-Bi2212 by temperature-dependent ARPES experiments as suggested in previous studies. On top of that, possible suppression of the nematicity in the superconducting state was also indicated. However, t}here are still unsolved issues regarding the pseudogap, that is, how the nematicity is related to the other proposed orders inside the pseudogap phase, e.g., CDW \cite{Ghiringhelli821,Chang2012Direct-obs,Comin1335,Le-Tacon2013Inelastic-,Comin390} and loop current order \cite{PhysRevLett.96.197001,PhysRevB.89.094523,PhysRevLett.118.097003,0953-8984-26-50-505701,2003.07556} which correspond to translational and time-reversal symmetry breaking, respectively, and how the $\vec{Q}=0$ nematic order opens pseudogap in the quasi-particle band dispersion. Therefore, further work is necessary to identify the nature and the origin of nematicity.

\section*{Acknowledgements}
Informative discussion with T. K. Lee, H. Yamase, T. Shibauchi and K. Ishida is gratefully acknowledged.  A part of this work was conducted at Advanced Characterization Nanotechnology Platform of the University of Tokyo, supported by "Nanotechnology Platform" of the Ministry of Education, Culture, Sports, Science and Technology (MEXT), Japan. ARPES experiments were performed at UVSOR (Proposal Nos. 28-813 and 29-821). This work was supported by KAKENHI Grant No. 15H02109 and 19K03741, and by “Program for Promoting Researches on the Supercomputer Fugaku” (Basic Science for Emergence and Functionality in Quantum Matter - Innovative Strongly-Correlated Electron Science by Integration of “Fugaku” and Frontier Experiments) from MEXT.

%

\end{document}



\title{Supplementary information for \\Nematicity in the pseudogap state of cuprate superconductors revealed by angle-resolved photoemission spectroscopy}

\author{S. Nakata}
\altaffiliation[e-mail: ]{S.Nakata@fkf.mpg.de}
\affiliation{Department of Physics, University of Tokyo, Bunkyo-ku, Tokyo 113-0033, Japan}
\author{M. Horio}
\affiliation{Department of Physics, University of Tokyo, Bunkyo-ku, Tokyo 113-0033, Japan}
\author{K. Koshiishi}
\affiliation{Department of Physics, University of Tokyo, Bunkyo-ku, Tokyo 113-0033, Japan}
\author{K. Hagiwara}
\affiliation{Department of Physics, University of Tokyo, Bunkyo-ku, Tokyo 113-0033, Japan}
\author{C. Lin}
\affiliation{Department of Physics, University of Tokyo, Bunkyo-ku, Tokyo 113-0033, Japan}
\author{M. Suzuki}
\affiliation{Department of Physics, University of Tokyo, Bunkyo-ku, Tokyo 113-0033, Japan}
\author{S. Ideta}
\affiliation{UVSOR Facility, Institute for Molecular Science, Okazaki 444-8585, Japan}
\author{K. Tanaka}
\affiliation{UVSOR Facility, Institute for Molecular Science, Okazaki 444-8585, Japan}
\author{D. Song}
\affiliation{National Institute of Advanced Science and Technology (AIST), Tsukuba 305-8568, Japan}
\author{Y. Yoshida}
\affiliation{National Institute of Advanced Science and Technology (AIST), Tsukuba 305-8568, Japan}
\author{H. Eisaki}
\affiliation{National Institute of Advanced Science and Technology (AIST), Tsukuba 305-8568, Japan}
\author{A. Fujimori}
\affiliation{Department of Physics, University of Tokyo, Bunkyo-ku, Tokyo 113-0033, Japan}
\affiliation{Department of Applied Physics, Waseda University, Shinjuku-ku, Tokyo 169-8555, Japan}
\date{\today}

\pacs{Valid PACS appear here}
\maketitle

\tableofcontents

\clearpage

\section*{Supplementary Note}

\subsection{Magnitude of strain estimated from the lattice parameters }

\begin{figure}[h]
\includegraphics[width = 15cm]{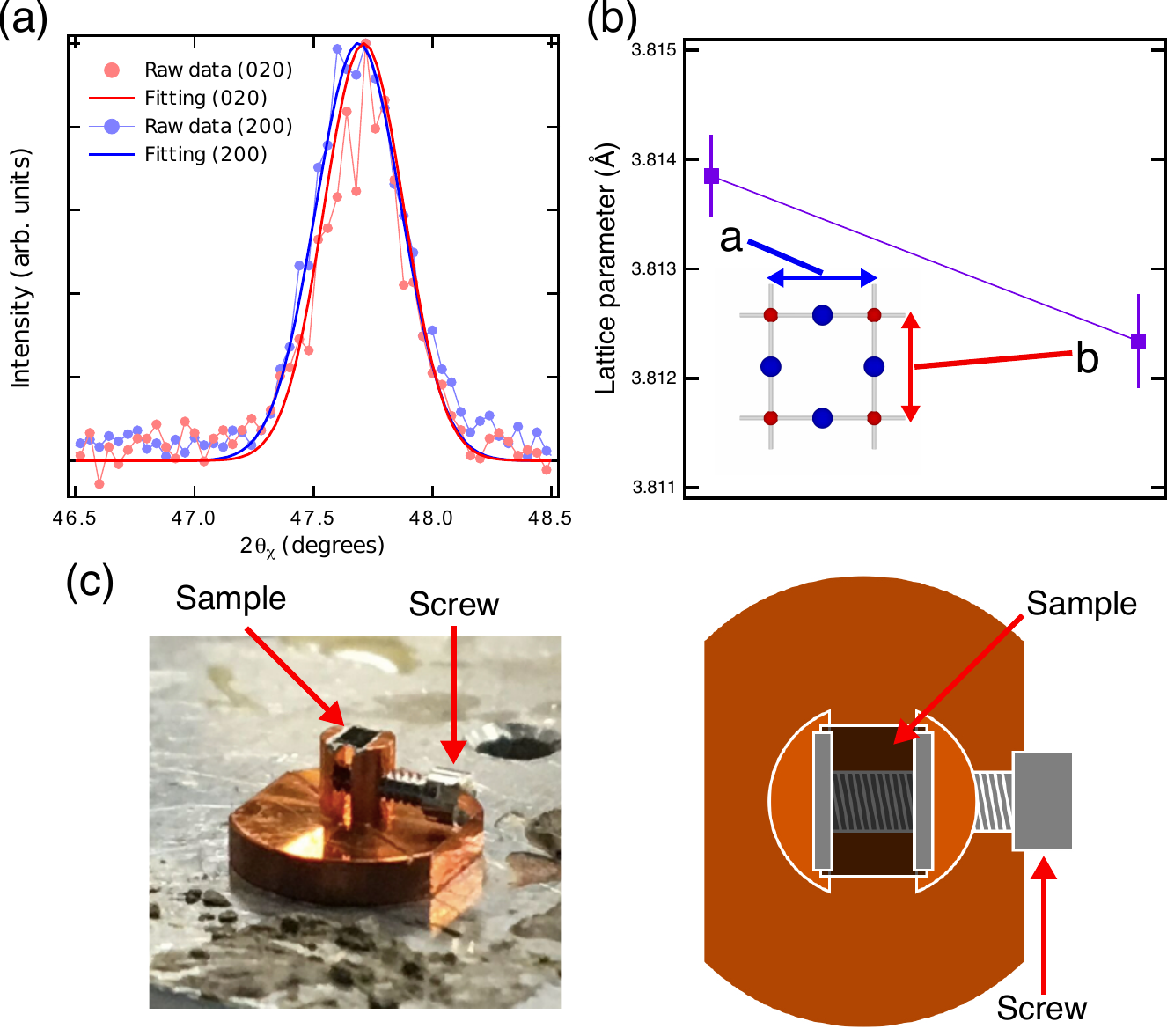}
\caption{\label{fig:S2} (a) X-ray diffraction pattern around the (200) and (020) peaks of Pb-Bi2212. Data were fitted using a single Gaussian. (b) Lattice parameters of $a$ and $b$ estimated from the Bragg peaks using Bragg's law.  The difference between $a$ and $b$ was less than 0.03 \%. (c) Sample holder designed to apply uniaxial tensile strain to the sample by tightening the stainless screw. The top view is shown on the right.}
\end{figure}

After the ARPES measurements, we carried out x-ray diffraction measurements in order to estimate the magnitude of the uniaxial tensile strain applied to the sample by the screw shown in Fig. \ref{fig:S2}(c). We estimated the strain from the anisotropy of the $a$ and $b$ lattice parameters because it is difficult to measure the strain directly because of the plasticity of the device. As shown in Fig. \ref{fig:S2}(a), we focused on the (200) and (020) Bragg peaks of Pb-Bi2212 and each peak was fitted using a single Gaussian. Figure \ref{fig:S2}(b) shows the $a$ and $b$ lattice parameters estimated to be as $a = 3.814$  \AA \, and $b = 3.812$ \AA. The anisotropy of the crystal structure is $(a-b)/(a+b) < 0.03$ \%. The anisotropy of the electronic structure  induced by the distortion $\delta_\text{crystal} = 7 \times (a-b)/(a+b) < 0.2$ \% expected from Eq. (\ref{eq:3.2.5}) is much smaller than the anisotropy parameter $\delta$ of order 1\% which we deduced from the ARPES measurements shown in Fig. 3. Therefore, the enhancement of the nematicity in the pseudogap state cannot be explained due merely to the anisotropy of the crystal structure.
A picture and a schematic figures of the strain device are shown in Fig \ref{fig:S2}(c).

\clearpage

\subsection{Band bottoms in the anti-nodal region in the pseudogap state}

\begin{figure}[h]
\begin{center}
\includegraphics[width=\textwidth/2]{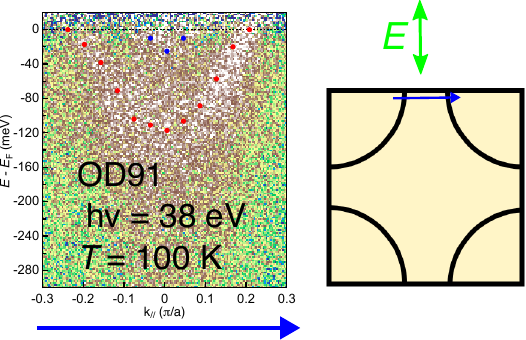}
\caption{Energy-momentum intensity map and EDC peaks corresponding to the bonding and anti-bonding bands for a cut crossing $\Vec{k}$ = $(0, \pi)$ in the pseudogap state of the overdoped ($T_\text{c}$ = 91 K) sample. The bottom of the bonding band is located at around -110 meV while the bottom of the anti-bonding band is barely (around -25 meV) below $E_\text{F}$. The right panel shows the momentum cuts of the energy dispersion and the polarization of the incident light.}
\label{fig:S1}
\end{center}
\end{figure}

The bottom energy of the bonding band at $(0, \pi)$ of the overdoped ($T_\text{c}$ = 91 K) sample studied in the present work was around -110 meV, which was estimated from the peak position of the corresponding EDCs as shown in Fig. \ref{fig:S1}. {The bottom of the antibonding band is also barely seen below $E_\text{F}$ ($\simeq$-25 meV).
Near the Fermi level, the MDCs are broad ($\sim0.3\pi/a$ FWHM compared to the antibonding-bonding separation of $\sim0.1\pi/a$), and therefore, it was difficult to assign separate Fermi surfaces of the anti-bonding and bonding bands (see details in supplementary information S5).} That is why the constant-energy surfaces estimated from peak positions of MDCs are not split into those of the anti-bonding and bonding bands, as shown in Fig. 2. On the other hand the bottoms of the anti-bonding and bonding bands are barely seen in the EDCs around $\Vec{k}$ = $(0, \pi)$ and $(\pi, 0)$ as shown in Fig. 1 of the main text. Those features are also seen  in previous ARPES studies \cite{Chen1099}.
Based on the estimates of the energies of the band bottoms, the line shape of the EDCs shown in Fig. 1 is discussed qualitatively in the main text.

\clearpage

\subsection{Energy distribution curves in the nordal region}

In the nematic state where the electronic structure becomes inequivalent between the $k_x$ and $k_y$ directions, one expects that the electronic structure around the node is less affected by the nematicity. In order to see if this is the case or not, we compare energy distribution curves (EDCs) around the nodes in the first and third quadrants of the first Brillouin zone at 125 K, the temperature at which the greatest anisotropy was observed between the $a$ and $b$ directions.

\begin{figure}[h]
\begin{center}
\includegraphics[width=\textwidth/2-1cm]{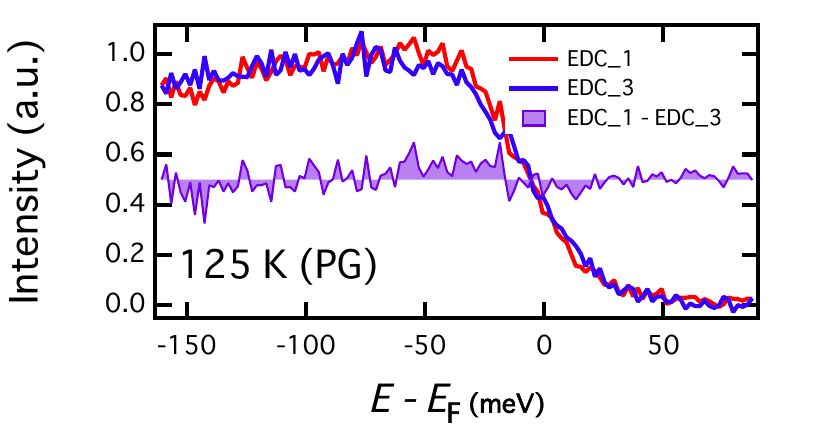}
\includegraphics[width=\textwidth/2-1cm]{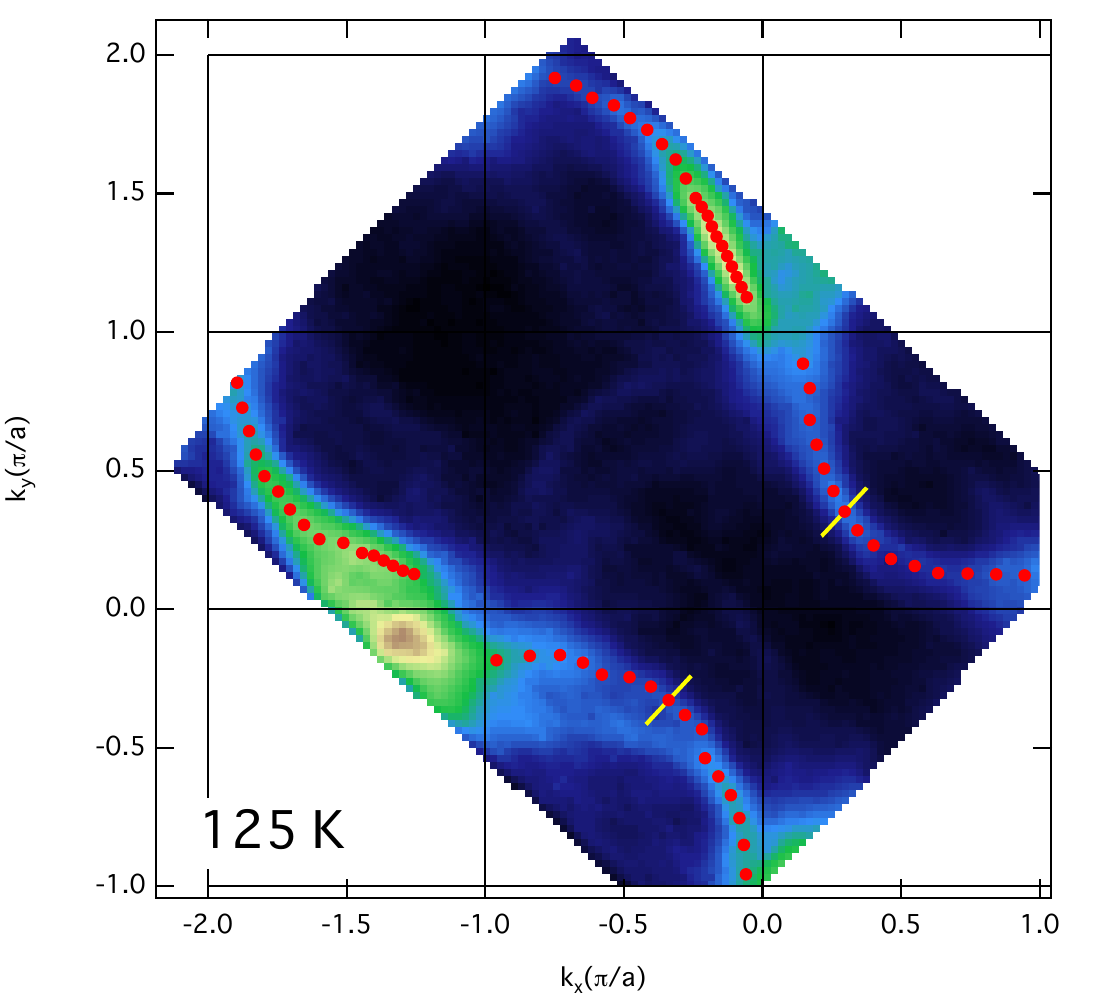}
\caption{(Left) Energy distribution curves (EDCs) around the nodes in the first and third quadrants of the first Brillouin zone at 125 K. The subtraction between the EDCs are also shown with the offset in the same manner as Fig. 1 in the main text. (Right) Constant energy surface at $E-E_\text{F} = -30$ meV at 125 K. The yellow lines indicate the integrated momentum ranges to obtain the EDCs in the left figure.}
\label{fig:SnodeEDC}
\end{center}
\end{figure}

As shown in Fig. \ref{fig:SnodeEDC}, the EDCs around the nodes are almost identical at the two different points in the first Brillouin zone in contrast to the EDCs at the X and Y points shown in the Fig. 1 (F).

\clearpage

\subsection{Matrix-element effect}

In vacuum, the vector potential $\vec{A}(\vec{r},t)$ satisfies the wave equation,
\begin{equation}
\nabla ^2 \vec{A}(\vec{r},t) = \frac{1}{c^2} \frac{\partial ^2 \vec{A}(\vec{r},t)}{\partial t^2}, \label{eq:2.5.1}
\end{equation}
and one of the solutions of this equation is
\begin{equation}
\vec{A}(\vec{r},t) = \vec{\varepsilon}e^{i(\vec{k}\cdot\vec{r}-\omega t)}, \label{eq:2.5.2}
\end{equation}
where $\vec{\varepsilon}$ denotes the polarization (the oscillating direction of the electric field), and $\vec{k}$ and $\omega$ are the wave vector and the frequency of light, respectively ($\omega = c k$). The probability of photoemission is given by
\begin{equation}
w_{\text{i} \rightarrow \text{f}} \propto \left|\braket{f|H'|i}\right|^2,\label{eq:2.5.3}
\end{equation}
where
\begin{equation}
H' = \frac{e}{m}\vec{A}(\vec{r})\cdot \vec{p} = \frac{e}{m}\vec{\varepsilon}e^{i\vec{k}\cdot\vec{r}}\cdot \vec{p},\label{eq:2.5.4}
\end{equation}
according to Fermi's golden rule. In ARPES measurements, we often use vacuum-ultra violet, whose photon energy is about 100 eV i.e., the wave length is about 100 \AA { }which is much longer than the distance between neighboring atoms in the material, so that one obtains
\begin{equation}
e^{i\vec{k}\cdot\vec{r}} = 1 + i\vec{k}\cdot\vec{r} + \cdots  \simeq 1  .\label{eq:2.5.5}
\end{equation}
Then the matrix element in Eq (\ref{eq:2.5.3}) can be expressed as
\begin{eqnarray}
\braket{f|H'|i} &\simeq& \braket{f|\frac{e}{m}\vec{\varepsilon}\cdot \vec{p}|i}\\ \label{eq:2.5.6}
&=& \frac{e}{i\hbar} \braket{f|\vec{\varepsilon}\cdot [\vec{x},H_0]|i}\\
&=& \frac{e}{i\hbar} \braket{f|\vec{\varepsilon}\cdot \vec{x}H_0^\dagger - H_0\vec{\varepsilon}\cdot \vec{x}|i}\\
&=& \frac{e}{i\hbar}(E_\text{f}-E_\text{i}) \braket{f|\vec{\varepsilon}\cdot \vec{x}|i},\label{eq:2.5.9}
\end{eqnarray}
using the Hermite property of $H_0$ and the commutation relationship
\begin{equation}
[\vec{x},H_0]=\frac{i\hbar \vec{p}}{m}.\label{eq:2.5.10}
\end{equation}
In summary, the transition probability can be approximated by
\begin{equation}
w_{\text{i} \rightarrow \text{f}} \propto \left|\braket{f|\vec{\varepsilon}\cdot \vec{x}|i}\right|^2,\label{eq:2.5.11}
\end{equation}
and the matrix element can be described by that of the dipole transition.\par
\begin{figure}[h]
\begin{center}
\includegraphics[width=13cm]{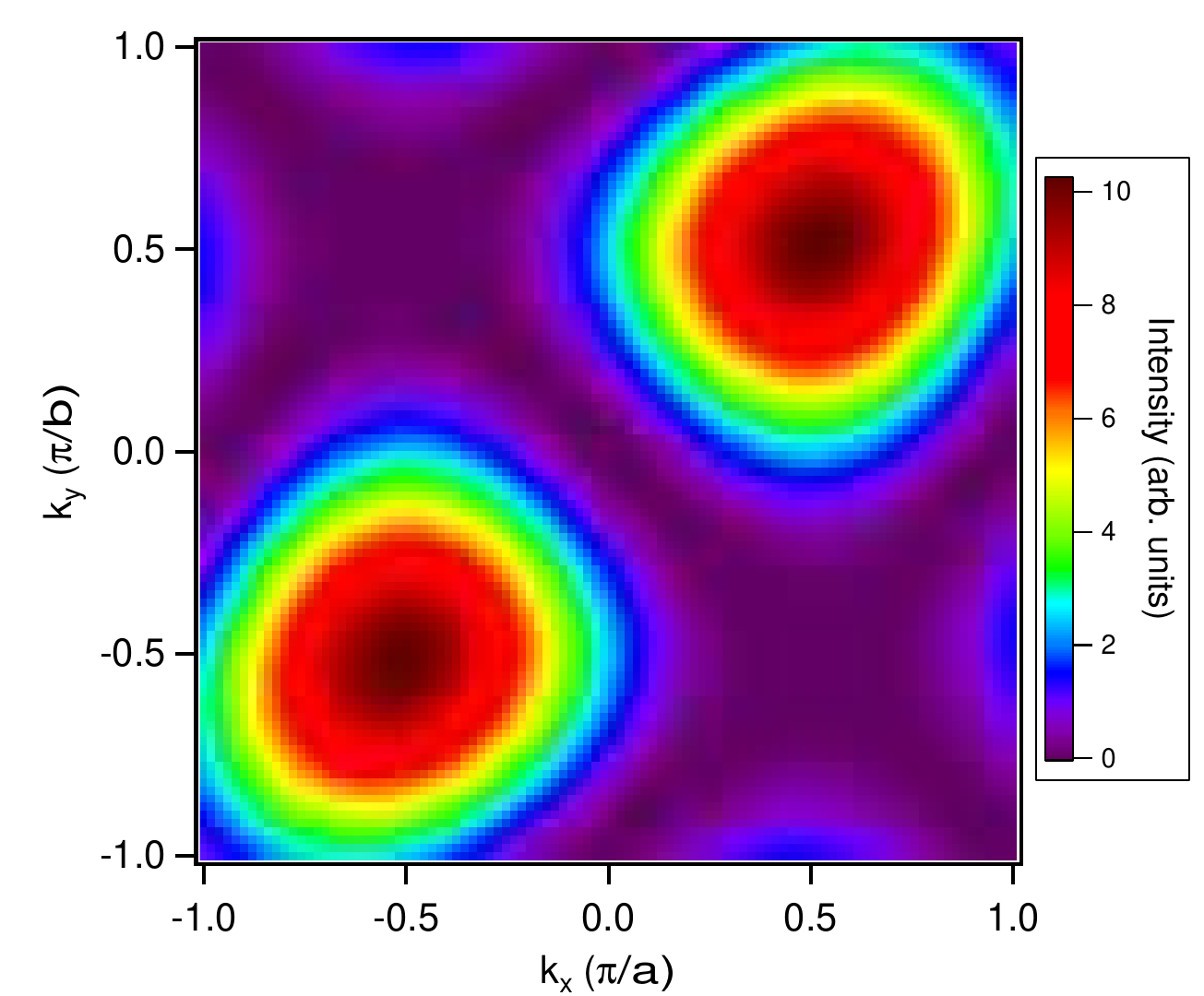}
\caption{Intensity plot in the first Brillouin zone for $\ket{i} = \ket{d_{x^2-y^2}}$, $\vec{E} \propto (+1,-1,0)$, and $\ket{f} = e^{i\vec{k}\cdot\vec{r}}$.}
\label{fig:2.5.1}
\end{center}
\end{figure}

The $k$-dependence of the intensity in $k$-space obtained by ARPES can be estimated by calculating the matrix element (Eq. (\ref{eq:2.5.11})) in the case that $\ket{i} = \ket{d_{x^2-y^2}}$, $\vec{E} \propto (+1,-1,0)$, which represents our experimental condition, and $\ket{f} = e^{i\vec{k}\cdot\vec{r}}$ was assumed for simplicity. Thus, we obtain
\begin{eqnarray}
|\braket{f|\vec{E}\cdot \vec{r}|i}|^2 &=& \left|\braket{e^{i\vec{k}\cdot\vec{r}}|\vec{E}\cdot \vec{r}|d_{x^2-y^2}}\right|^2\\
&\propto& \left|\int d^2\vec{r} (x^2-y^2)(x-y)\cos(k_xx+k_yy)\right|^2\\
&& +\left|\int d^2\vec{r} (x^2-y^2)(x-y)\sin(k_xx+k_yy)\right|^2.
\end{eqnarray}
Figure \ref{fig:2.5.1} is the result of the $k$-dependence of the transition probability in the first Brillouin zone. 

\clearpage

\subsection{Raw data of momentum distribution curves}

Here we show raw data of momentum distribution curves (MDCs) at all the measurement temperatures. As noted in the main text, each MDC has been integrated within $\pm$ 10 meV of $E-E_\text{F} = -30$ meV. The MDC at each tilt angle was fitted with Lorentzian functions plus a linear background around the peaks. The fitted range was chosen arbitrarily. The constant energy surface at $E-E_\text{F} = -30$ meV in Fig. 2 of the main text has been derived from the peak positions of the Lorentzian functions. For the analysis using the tight-binding model to estimate the nematicity plotted in Fig. 3, we have neglected data points which were seen due to the diffraction replica or arouse from the tilted part of the  cleaved surface in the heating series. Note that the intensity of the constant energy surface in the second quadrant is suppressed due to the matrix-element effect. This makes it difficult to compare the change in the band dispersion between near the X point and the Y point in a single MDC including both X and Y points. However, as explained in the main text, our experimental geometry is the only possible configuration that guarantees the symmetry with respect to $k_y = - k_x$ and this is why we had to resort to the simultaneous fitting of all the MDCs utilizing the tight-binding model.

\begin{figure}[h]
\begin{center}
\includegraphics[height=\textheight/3-1cm]{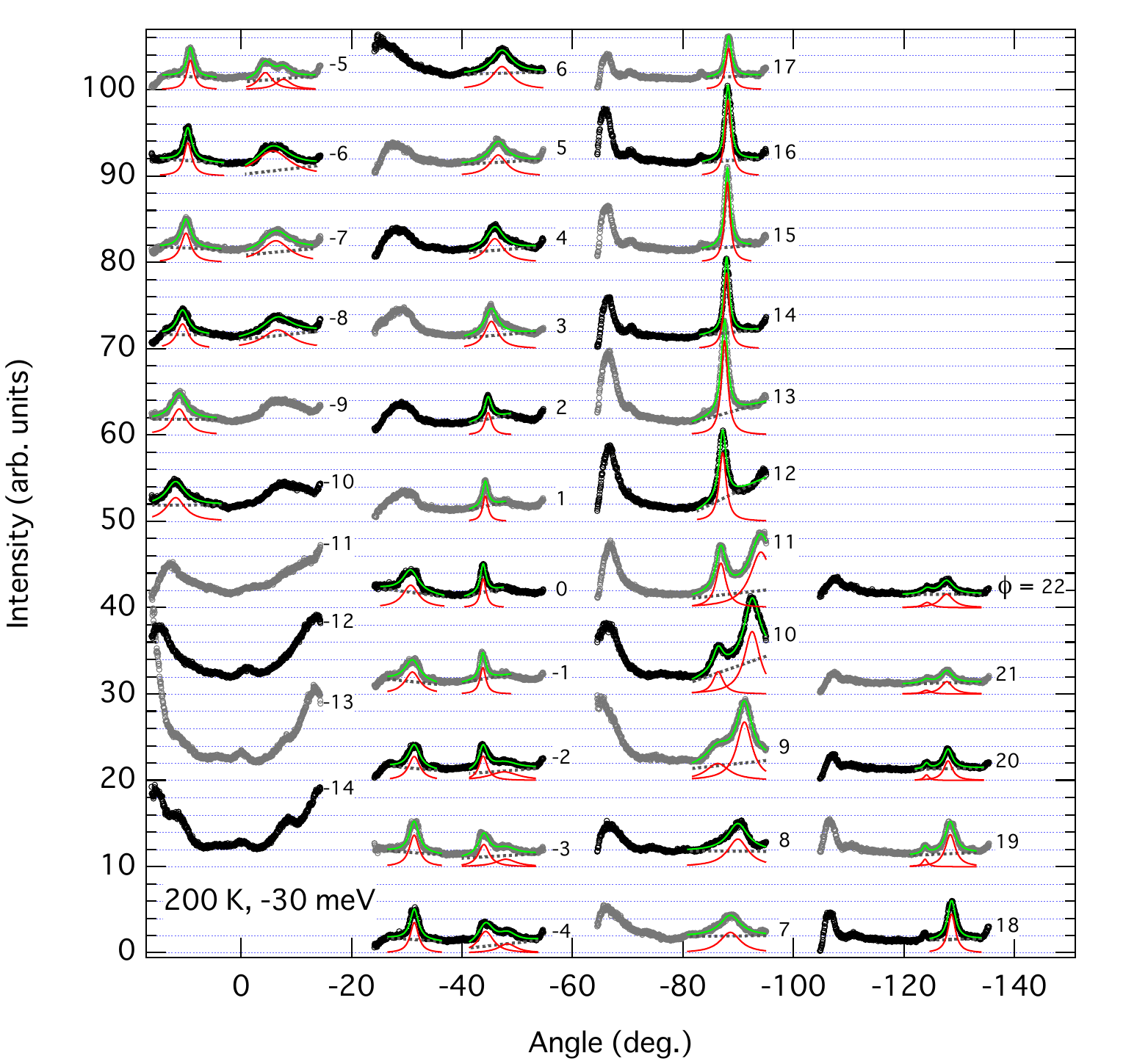}
\includegraphics[height=\textheight/3-1cm]{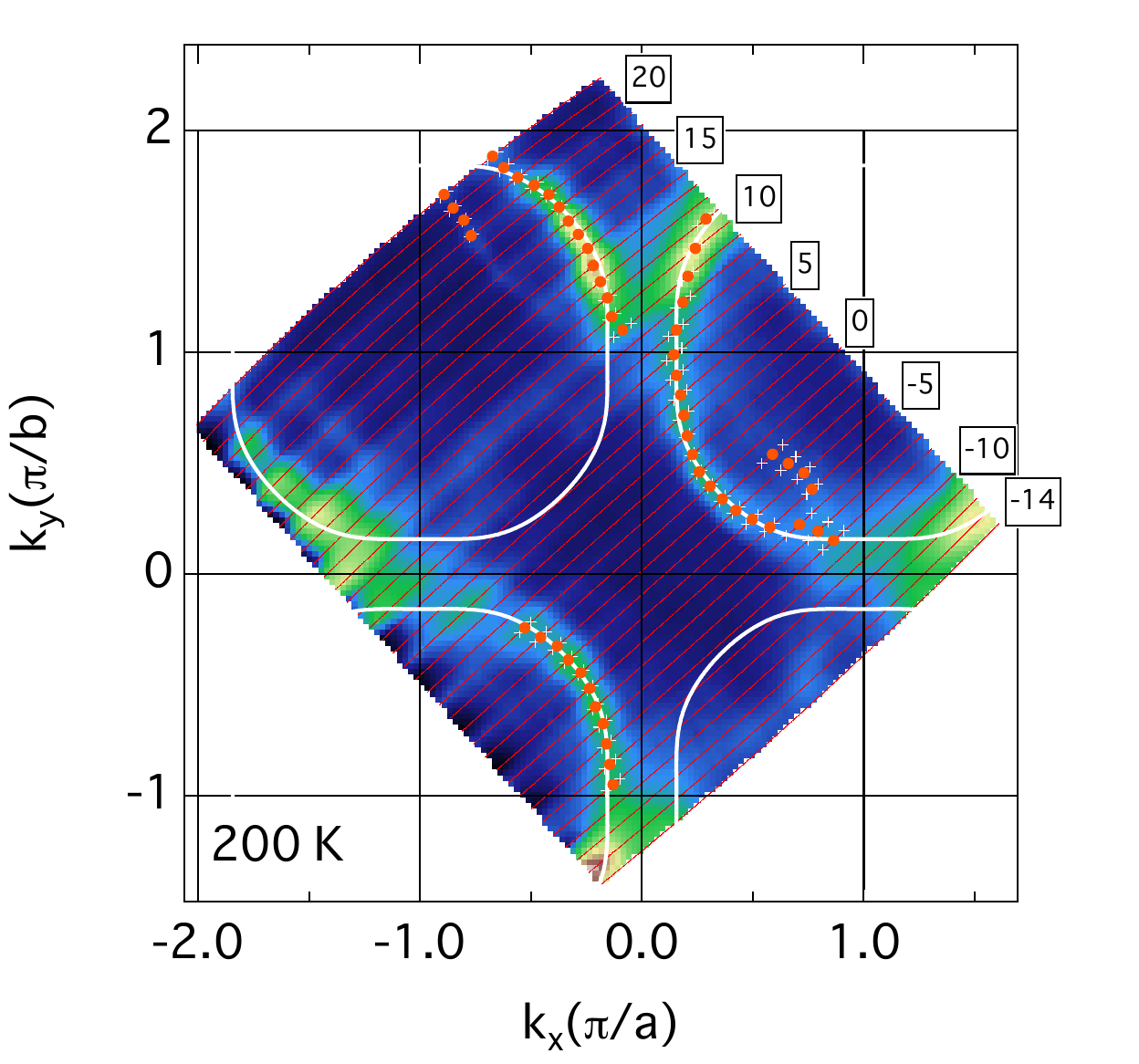}
\caption{(Left) MDC data at 200 K (black and gray curves). Red curves are Lorentzian functions, dotted gray lines are linear background, and green curves are the sum of them. Numbers next to each MDC indicates the tilt angle of the sample. For clarity, the data are shown with offset. (Right) Constant energy-surface mapping at $E-E_\text{F} = -30$ meV at 200 K. {Numbers next to MDC cuts indicates the tilt angle of the sample, corresponding to the ones in the left figure.} Orange points are the peak positions of the Lorentzian functions. Two white crosses around the orange point represent a quarter of the HWHM of the Lorentzian. White curves are fitted curves of the tight-binding model also shown in Fig. 2 of the main text.}
\label{fig:MDCs200K}
\end{center}
\end{figure}

\begin{figure}[h]
\begin{center}
\includegraphics[height=\textheight/3-1cm]{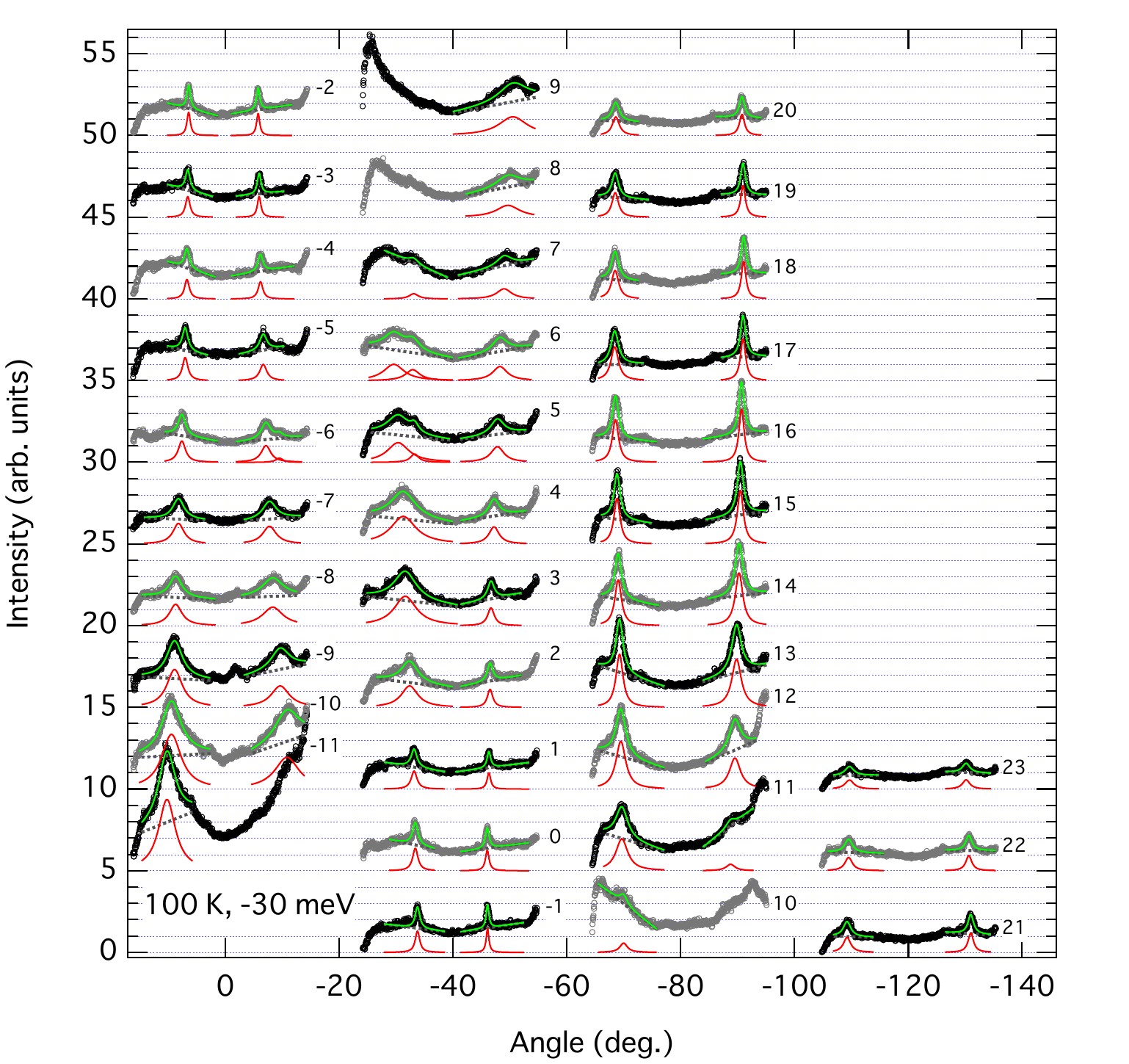}
\includegraphics[height=\textheight/3-1cm]{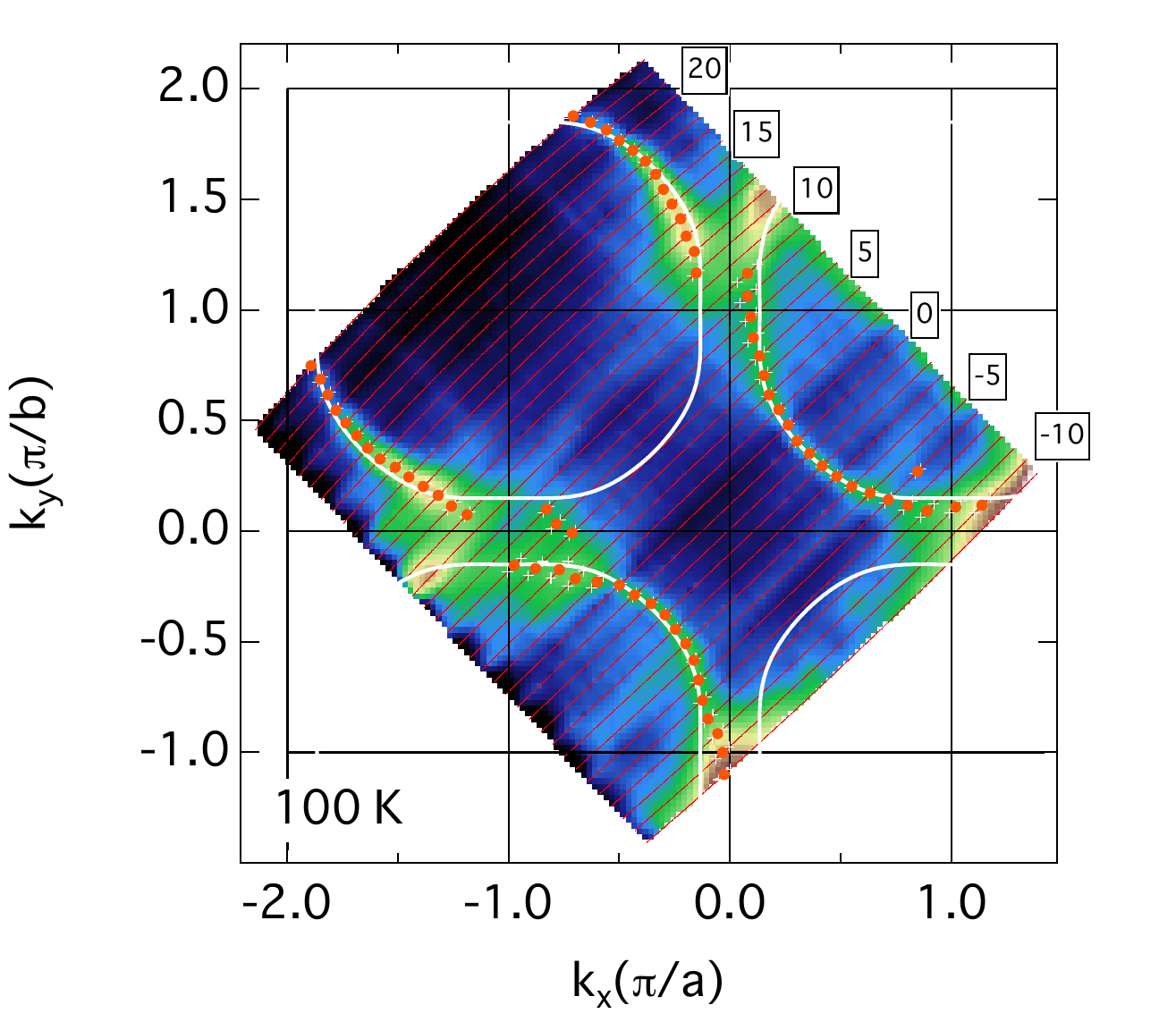}
\caption{The same as Fig. \ref{fig:MDCs200K} except that the data were taken at 100 K.}
\label{fig:MDCs100K}
\end{center}
\end{figure}

\begin{figure}[h]
\begin{center}
\includegraphics[height=\textheight/3-1cm]{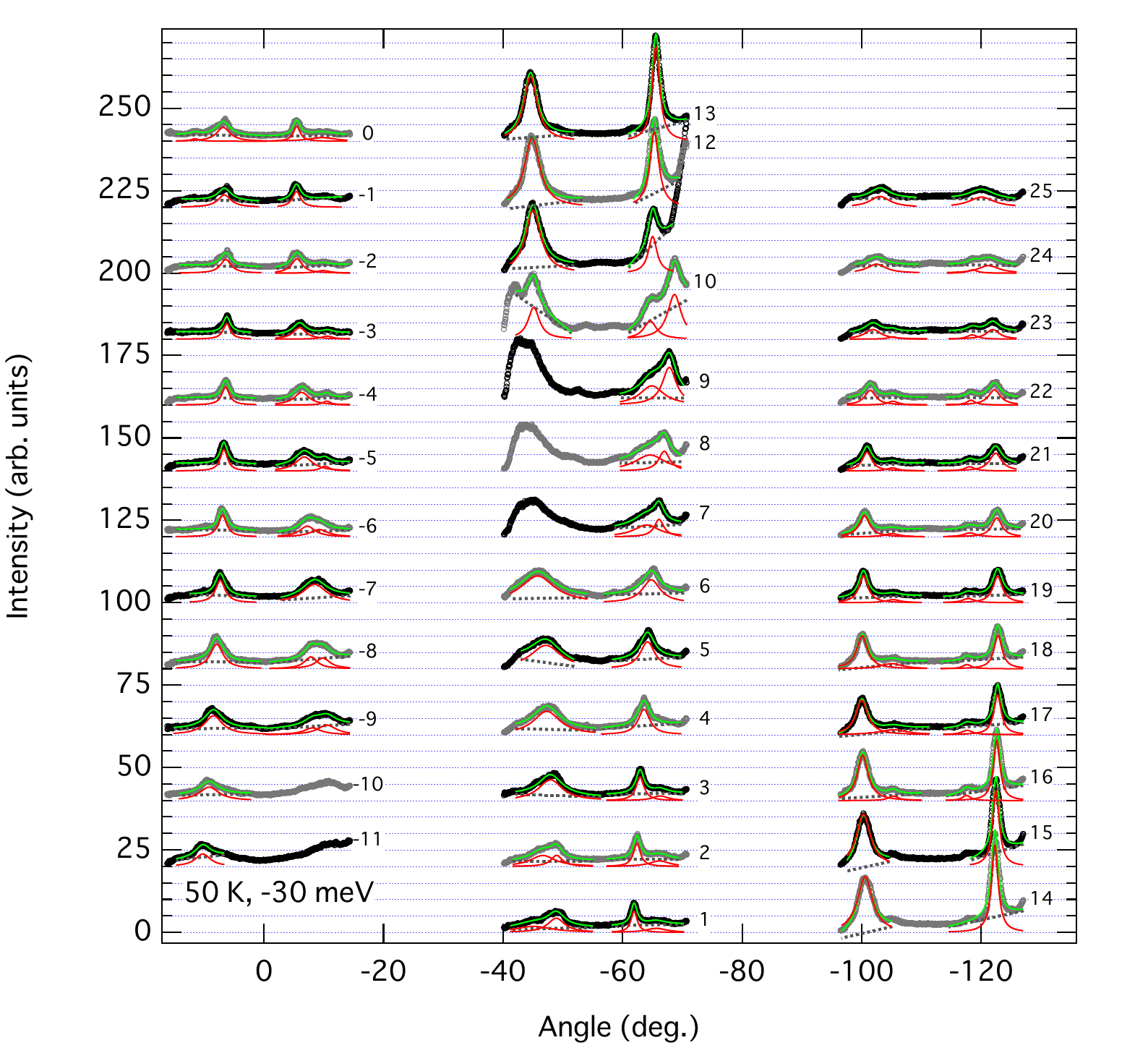}
\includegraphics[height=\textheight/3-1cm]{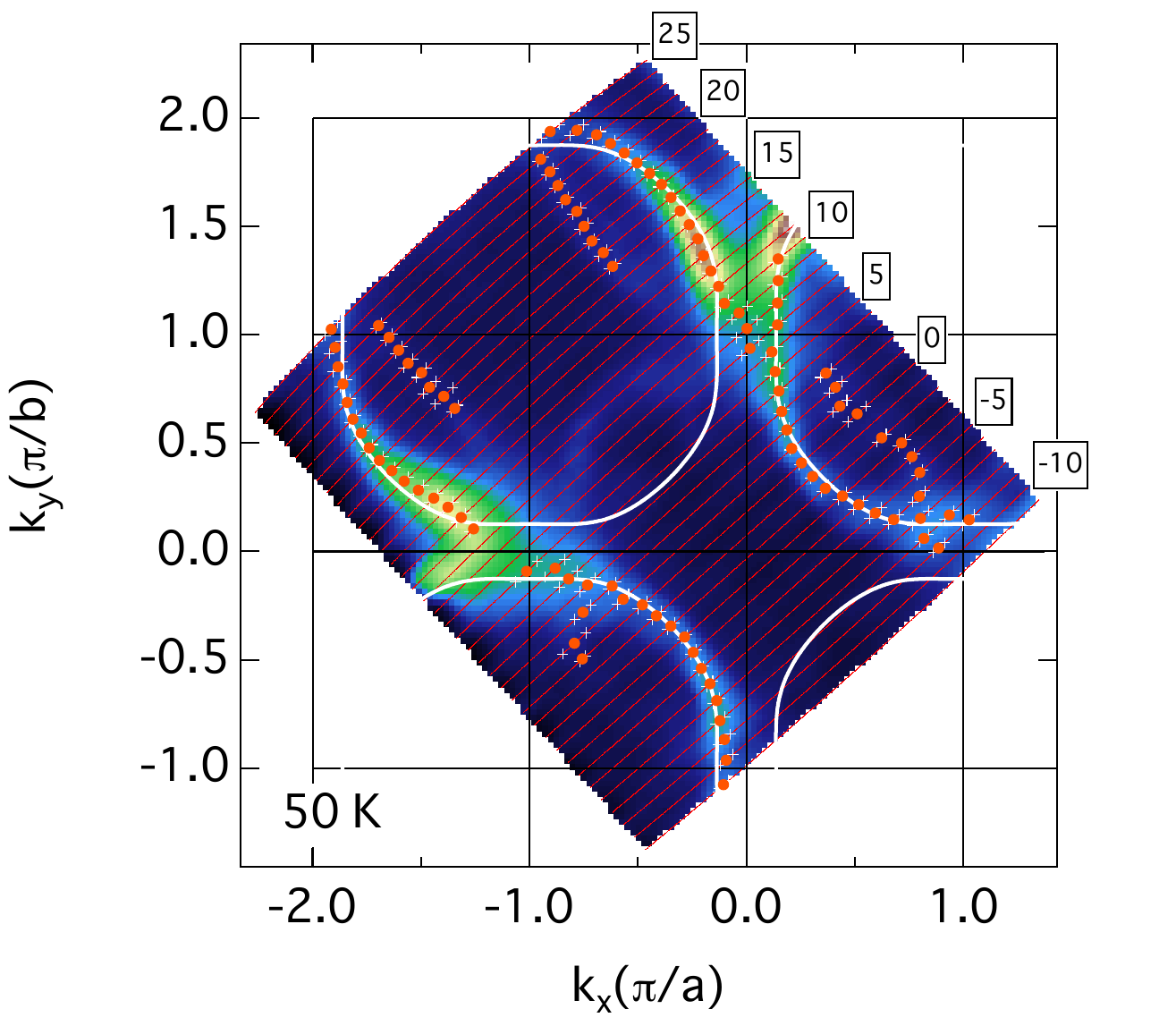}
\caption{The same as Fig. \ref{fig:MDCs200K} except that the data were taken at 50 K.}
\label{fig:MDCs50K}
\end{center}
\end{figure}

\begin{figure}[h]
\begin{center}
\includegraphics[height=\textheight/3-1cm]{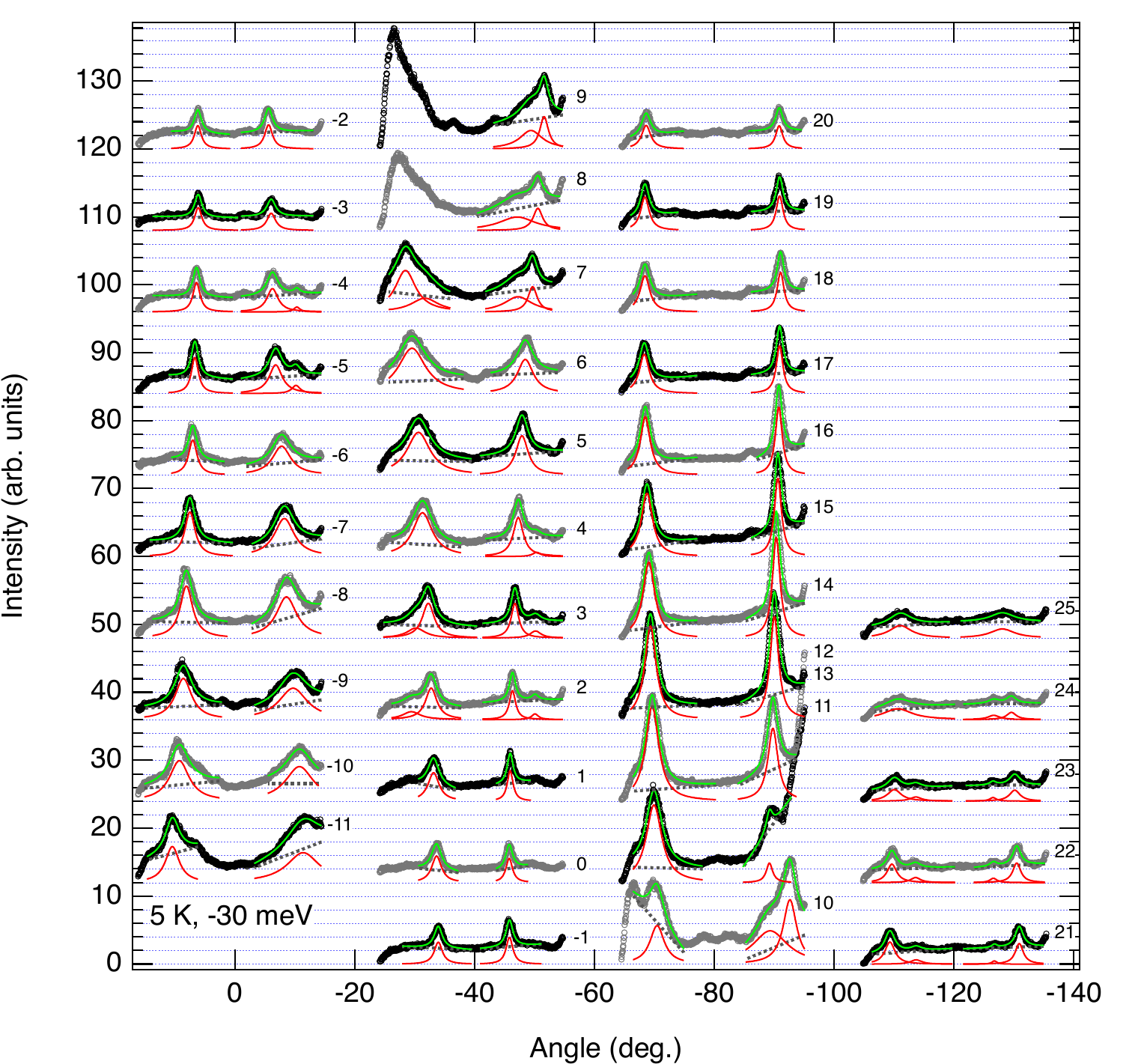}
\includegraphics[height=\textheight/3-1cm]{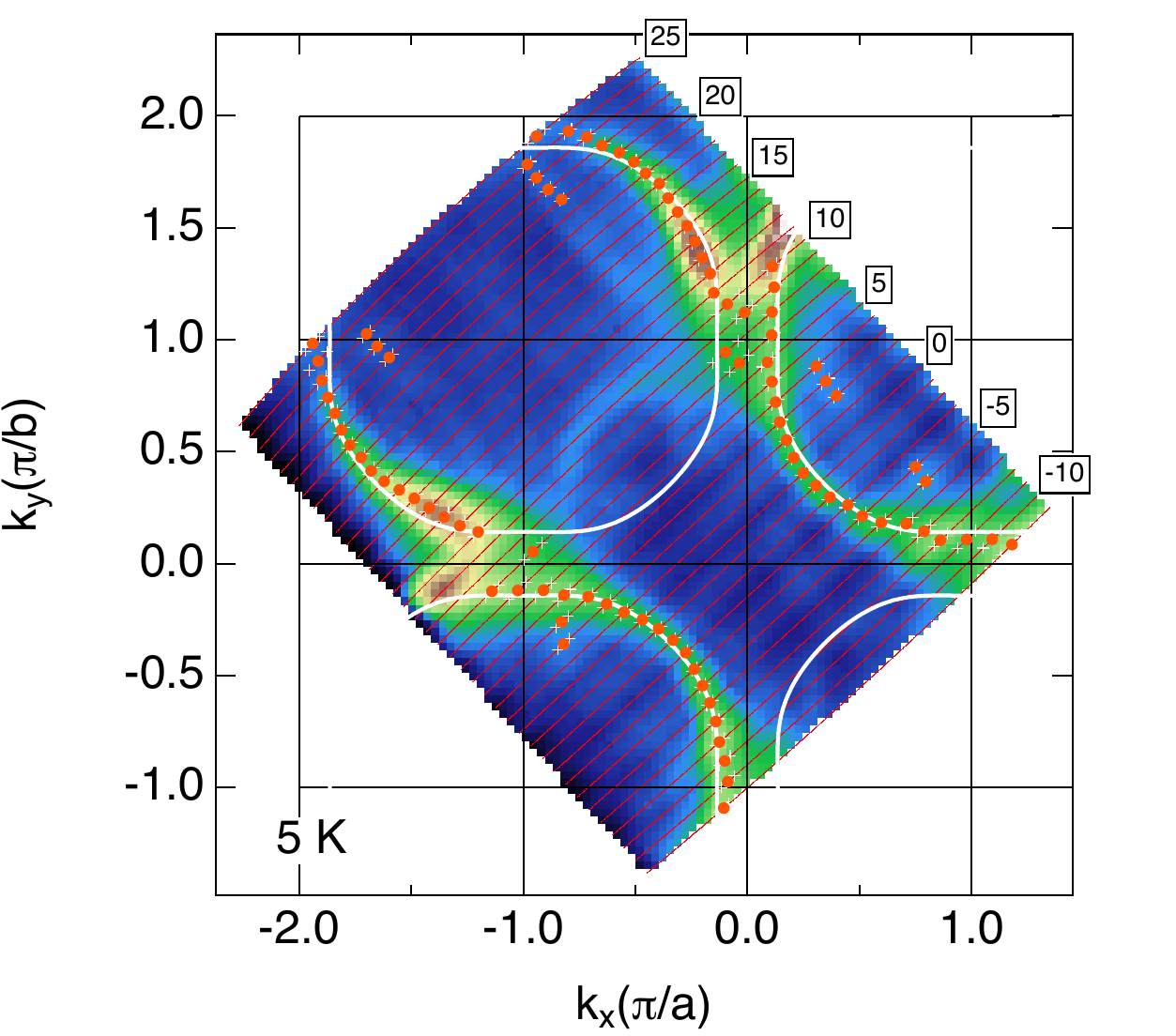}
\caption{The same as Fig. \ref{fig:MDCs200K} except that the data were taken at 5 K.}
\label{fig:MDCs5K}
\end{center}
\end{figure}

\begin{figure}[h]
\begin{center}
\includegraphics[height=\textheight/3-1cm]{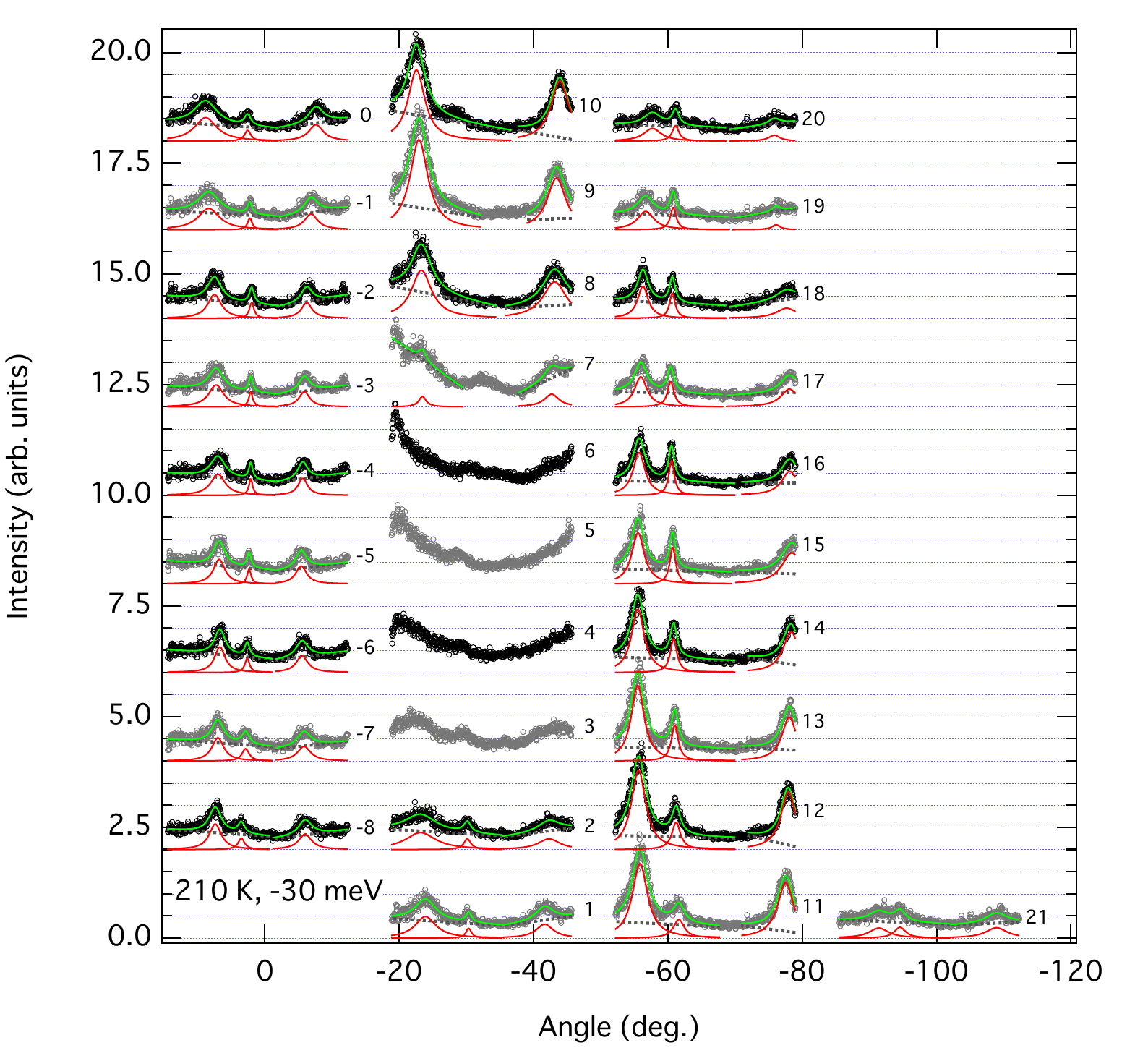}
\includegraphics[height=\textheight/3-1cm]{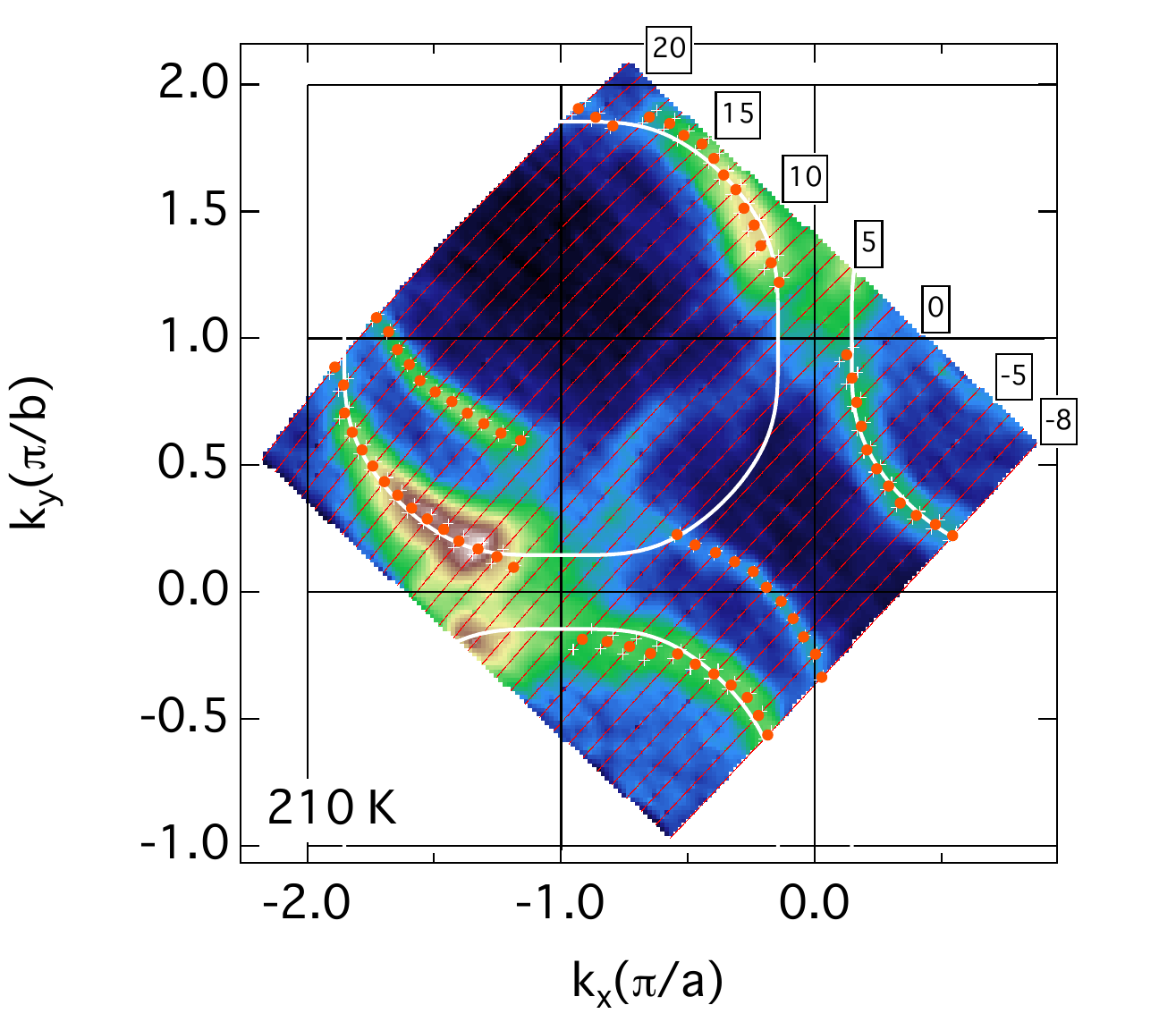}
\caption{The same as Fig. \ref{fig:MDCs200K} except that the data were taken at 210 K. Note that in the heating series (210, 125, and 25 K), MDC peaks from a possible tilted surface are seen unlike those in the cooling series (200, 100, 50, and 5 K). However, those features could be unambiguously separated from the intrinsic MDC peaks as described in Supplementary information S5, and thereby do not affect the tight-binding fit.}
\label{fig:MDCs210K}
\end{center}
\end{figure}

\begin{figure}[h]
\begin{center}
\includegraphics[height=\textheight/3-1cm]{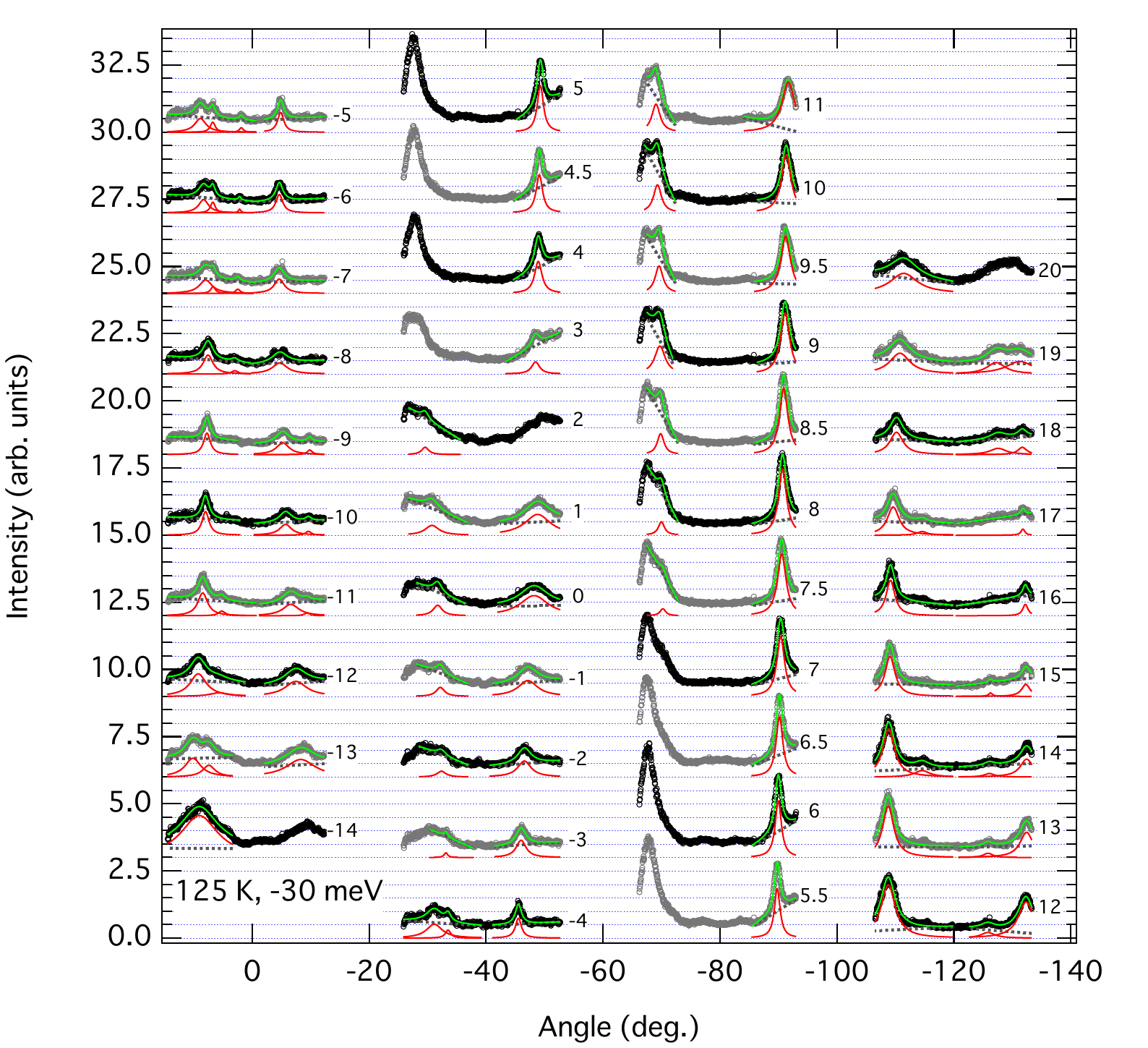}
\includegraphics[height=\textheight/3-1cm]{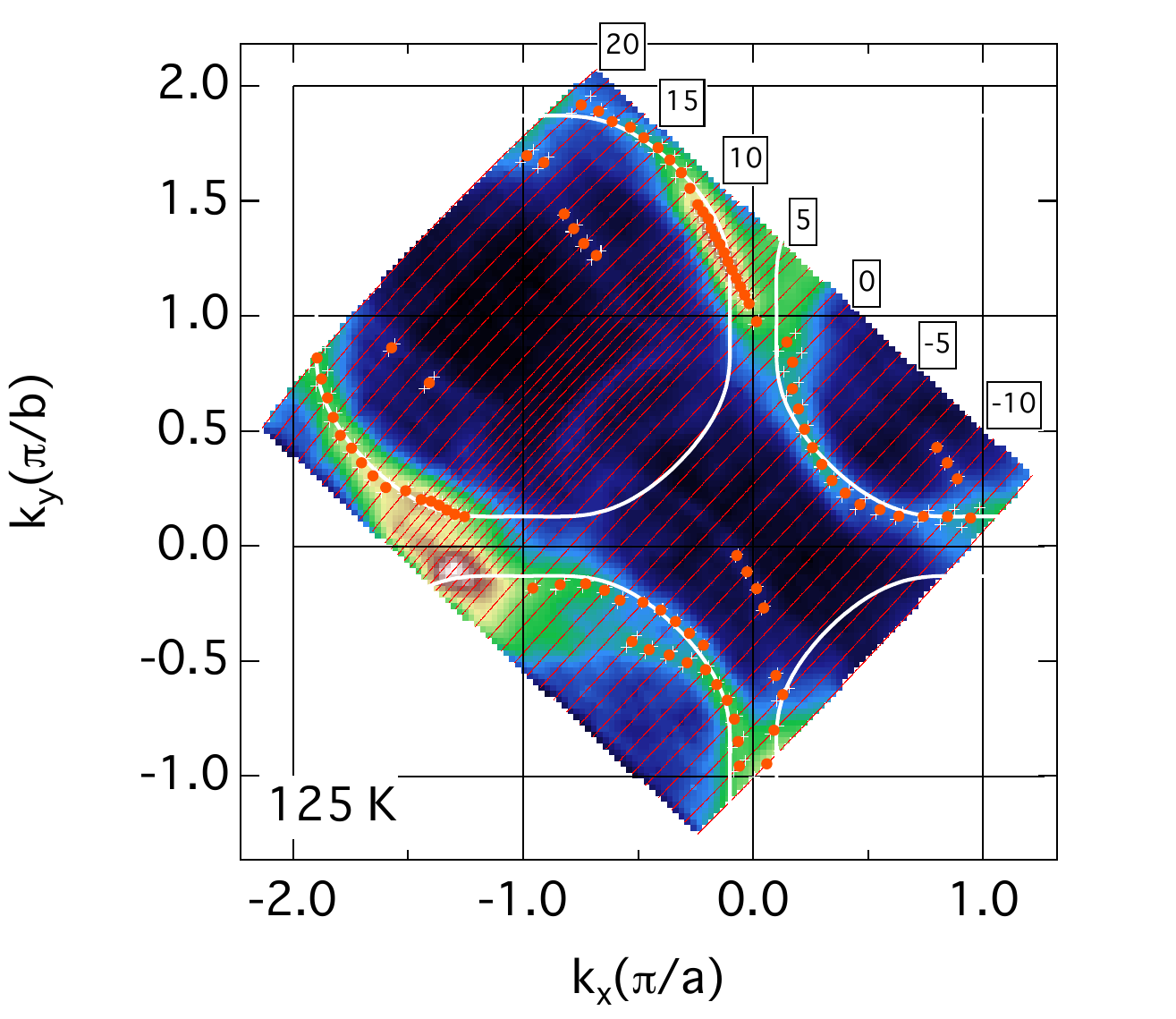}
\caption{The same as Fig. \ref{fig:MDCs210K} except that the data were taken at 125 K.}
\label{fig:MDCs5K}
\end{center}
\end{figure}

\begin{figure}[h]
\begin{center}
\includegraphics[height=\textheight/3-1cm]{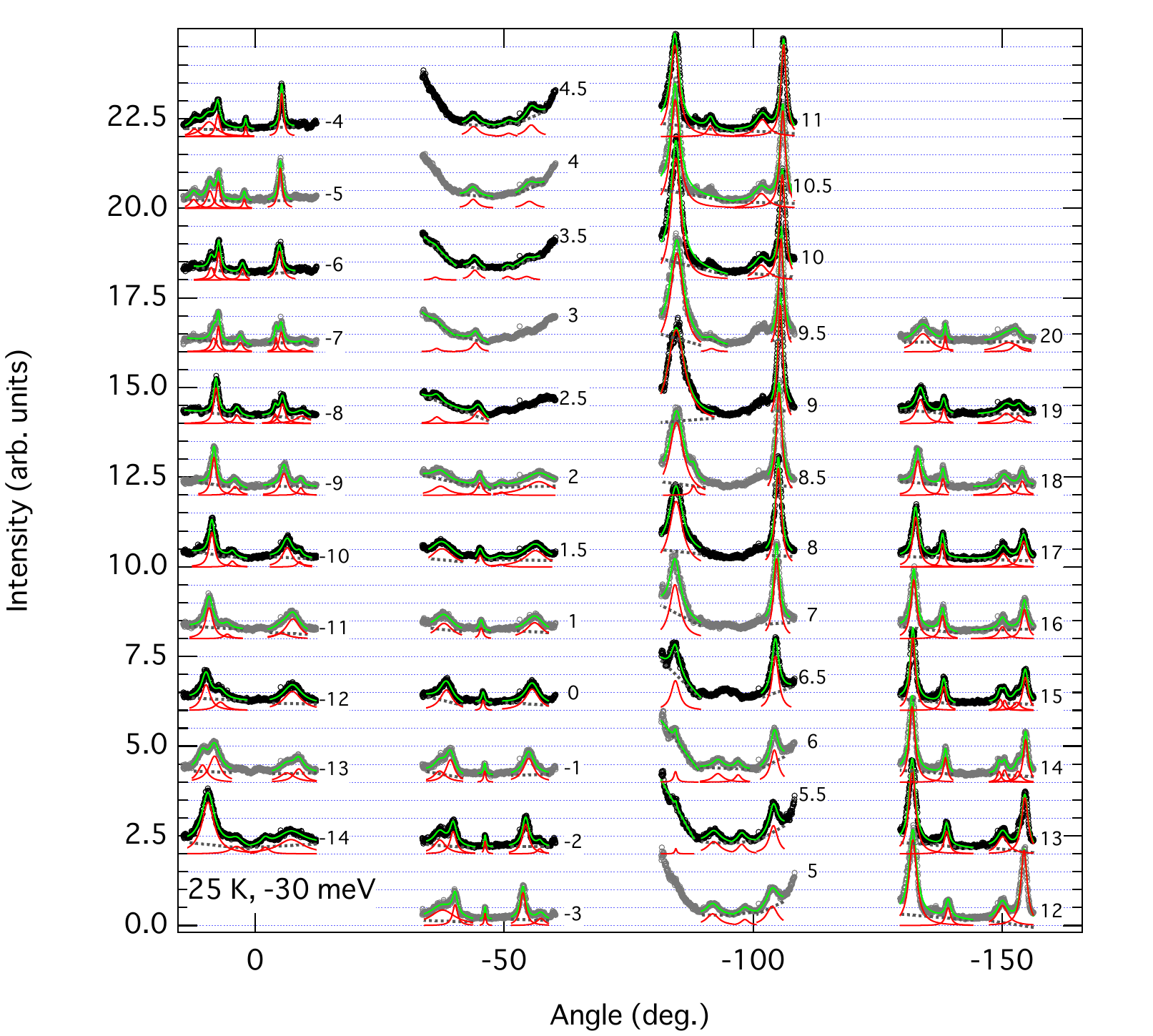}
\includegraphics[height=\textheight/3-1cm]{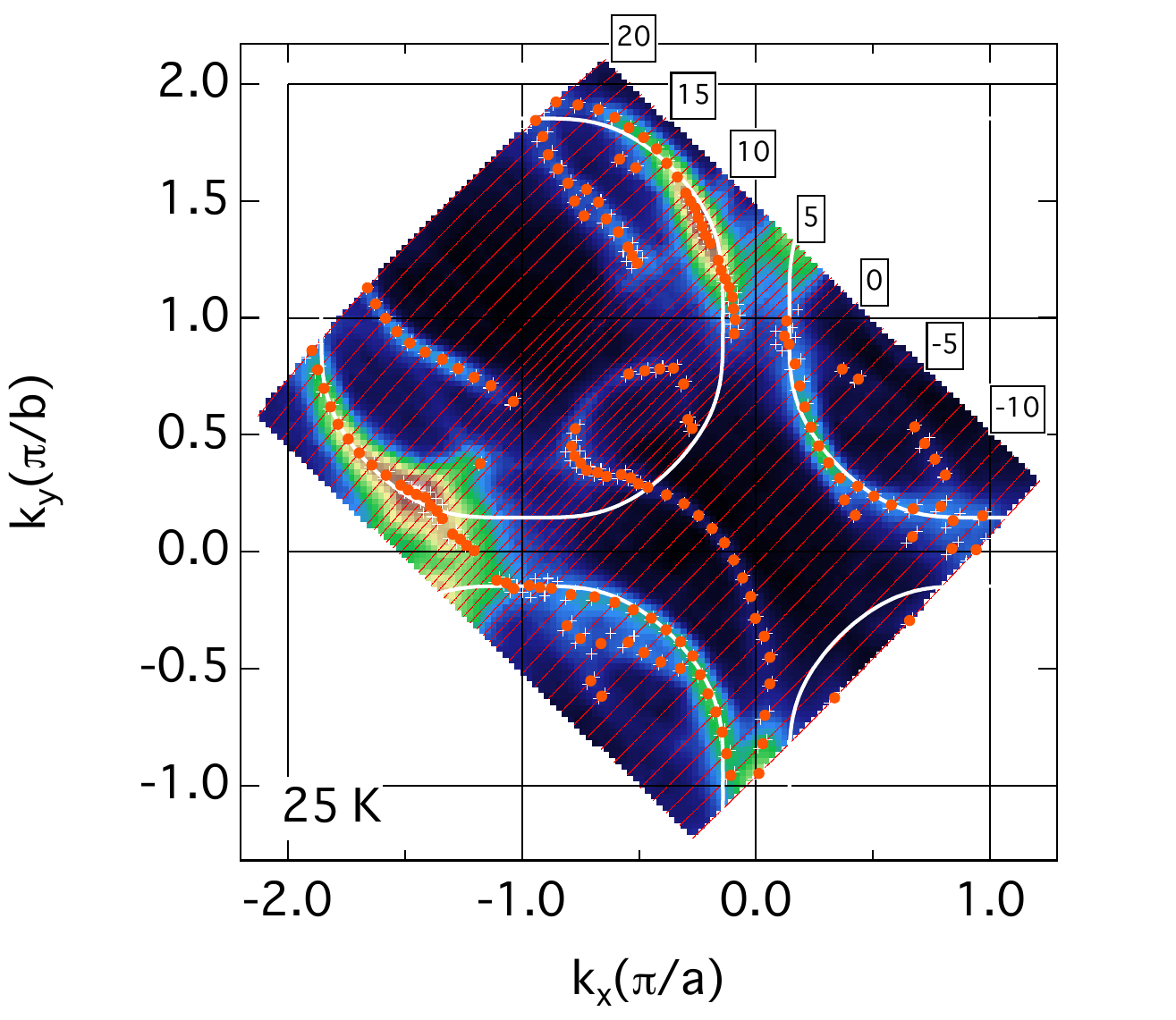}
\caption{The same as Fig. \ref{fig:MDCs210K} except that the data were taken at 25 K.}
\label{fig:MDCs5K}
\end{center}
\end{figure}

\clearpage

\subsection{Diffraction replicas}

As seen in the raw MDC data, we have observed some additional features in the constant-energy-surface mapping on top of the intrinsic feature analyzed in the main text. In the cooling series (200 K, 100 K, 50 K, and 5 K), so-called shadow bands, replicas of the main Fermi surface shifted by $\vec{k} = (\pi,\pi)$ of the structural origin \cite{PhysRevB.69.220505}, are seen. On top of that, the replicas due to a tilted part of the cleaved surface are also seen in the heating series (25K, 125 K, and 210 K). The shift is estimated to be $\vec{k} = (0.08\pi,0.25\pi)$. Some of those replicas, particularly of the $(\pi, \pi)$ replicas, intersect with the main feature in the antinodarl region, where MDC peak positions are crucial to study the distortion of the constant-energy surface representing the nematicity. Therefore, when the intensities of such extrinsic features are not negligibly small, the peak positions of MDCs may be deviated from the intrinsic ones. We have excluded such data points for analyzing the anisotropy of the constant-energy surfaces to eliminate the effect of extrinsic features.

\begin{figure}[h]
\begin{center}
\includegraphics[width=\textwidth/2-1cm]{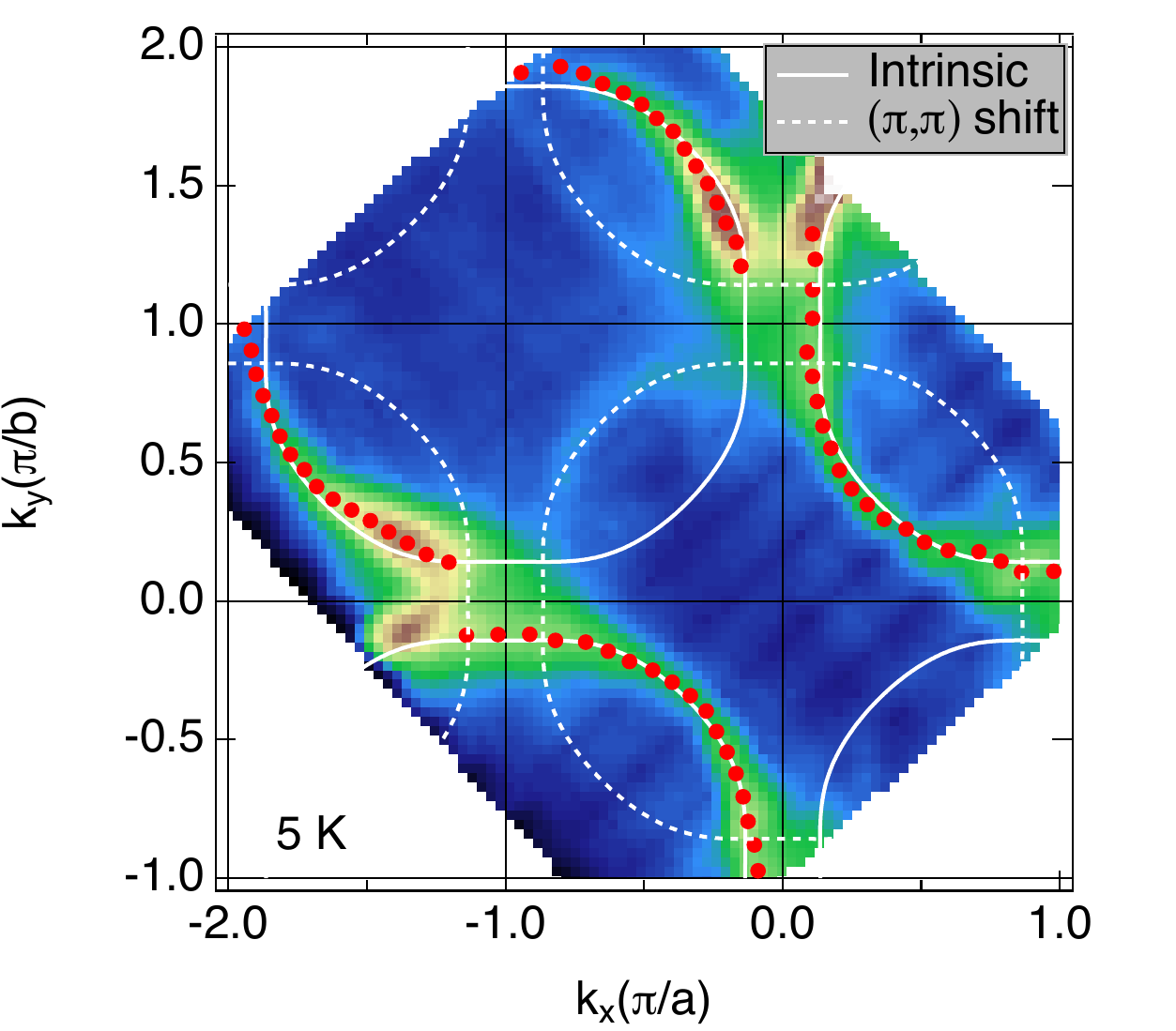}
\includegraphics[width=\textwidth/2-1cm]{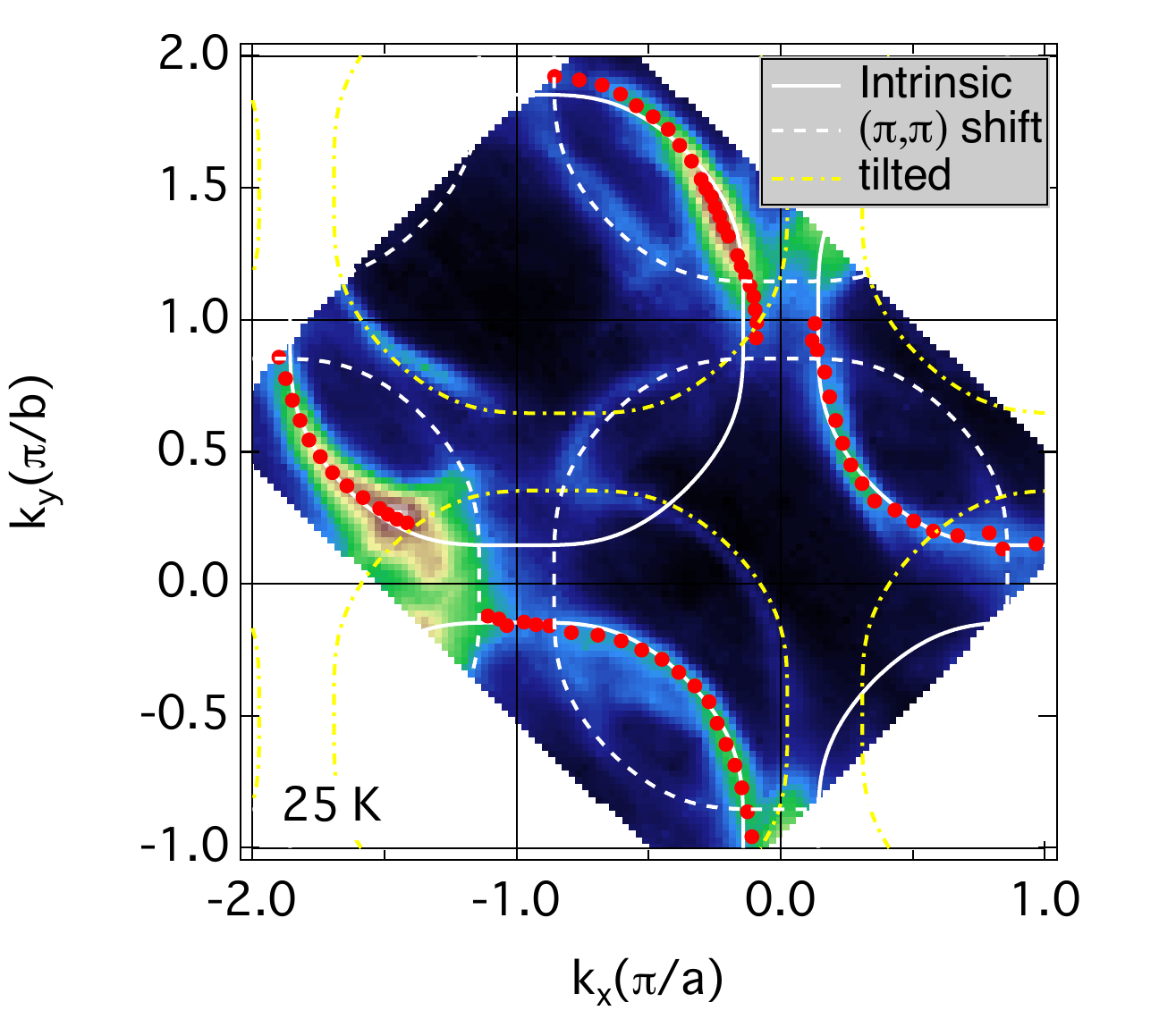}
\caption{(Left) The same as Fig. 2 (d) in the main text except for the indication of the diffraction replica or the so-called shadow band (dotted white curves). (Right) The same as Fig. 2 (g) in the main text except for the indication of the diffraction replicas due to the shadow band (dotted white curves) and another replica arising from a tilted part of the cleaved surface (dot-dashed yellow curves).}
\label{fig:MDCs5K}
\end{center}
\end{figure}

\clearpage

\subsection{Fitting procedure using the tight-binding model}

Fitting procedure to estimate the anisotropy of the constant energy surface shown in Fig. 3 is  as follows. First, the experimental data points which represent the constant energy surfaces at 200 K and 210 K, the highest temperature in the cooling and heating series, respectively, were fitted using the tight binding model Eq. (2).  Since the actual sample angle, i.e., the small misalignment of the sample, with respect to the sample stage cannot be determined with sufficient accuracy, we chose the {the misalignment angles between the sample and the sample stage}  $\theta_0, \phi_0, \omega_0$ for the rotation around the vertical axis, the tilt, and the azimuth rotation (see Fig. S13), respectively, such that the difference between the data points and fitted curve was minimized. At the highest temperatures in both measurement series (200 K and 210 K), only $t''/t'$ was fixed with the constraint: $-0.51 < t''/t' <-0.5$. {Note that, despite the fact that we did not assume $\delta = 0$ in the procedure to define the sample angles, the anisotropy parameter $\delta$ converged to almost zero.} At lower temperatures, $t'/t$ was also fixed within $\pm$ 1 \% of $t'/t$ at the highest temperatures. The {misalignment angles}  at lower temperatures were defined with the same method as the one at the highest temperatures{, however, using Eq. (1), i.e.,  imposing $\delta = 0$, instead of Eq. (2) where non-zero $\delta$ is allowed. The aim of this further constraint is to avoid the sample angles and the parameter $\delta$ from being trapped at a local minimum of the fitting error}. After defining the sample angles at each temperature, the data points were fitted with Eq. (2), i.e., $\delta$ was allowed to vary.

\begin{table}[h]
\centering
\begin{tabular}{c|ccc|cccc|c}
\hline
$T$ (K) & $\theta_0$ & $\phi_0$ & $\omega_0$ & $t'/t$ & $t''/t'$ &$\varepsilon_0/t$ & $\delta$ (\%) & FS area\\\hline
\hline 
5 & 0.3 & 1.65 & -2.65 & -0.28 & -0.5 & 1.26 & 0.49 & 1.23\\
50 & 0.3 & 1.7 & -3.25 & -0.28 & -0.5 & 1.30 & -0.72 &1.25\\
100 & 0.55 & 1.1 & -1.55 & -0.28 & -0.5 & 1.25 & 1.18 &1.22\\
200 & 0.7 & 2.15 & -3.25 & -0.28 & -0.5 & 1.18 & 0.18 &1.19\\\hline 
25 & 1.05 & 5.55 & 0.85 & -0.26 & -0.5 & 1.18 & 0.33 &1.20 \\
125 & 0.9 & 5.3 & 2.0 & -0.27 & -0.5 & 1.35 & 1.68 & 1.28 \\
210 & 0.65 & 4.9 & 2.15 & -0.27 & -0.5 & 1.18 & 0.03 &1.20 \\\hline
\end{tabular}
\caption{Misalignment sample angles ($\theta_0, \phi_0, \omega_0$) and best-fit parameter values for the tight binding model ($t'/t, t''/t', \varepsilon_0/t, \delta$). Fermi surface (FS) area is estimated with the tight-binding parameters. The FS areas are given in units of $2\pi^2$. The Fermi surface area slightly changes with temperature because the bonding and antibonding band intensities may change with nematicity.}
\end{table}

\begin{figure}[h]
\centering
\includegraphics[width = 8cm]{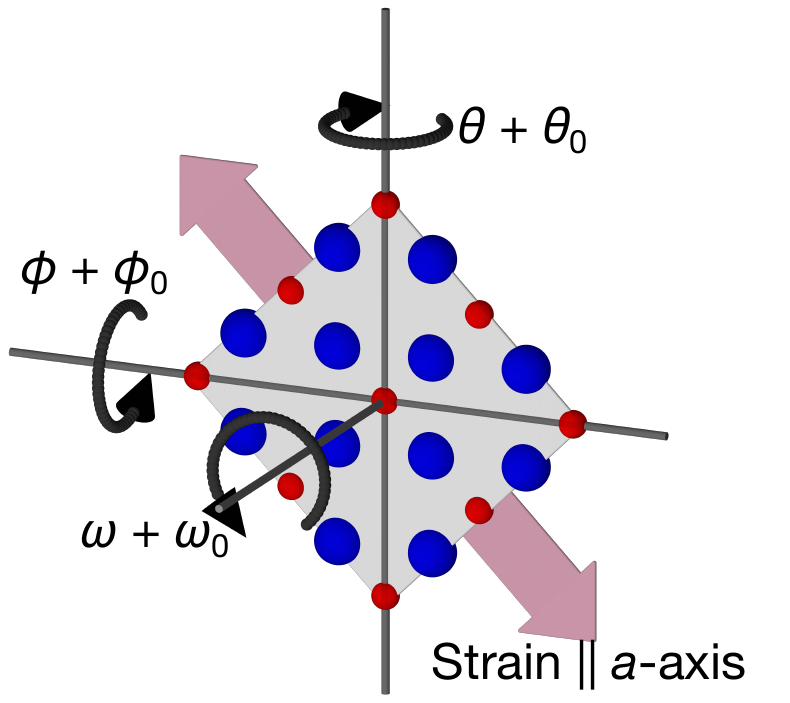}
\caption{Sample angles with respect to the sample stage. $(\theta, \phi, \omega)$ are angles controlled by the goniometer of the instrument. $(\theta_0, \phi_0, \omega_0)$ represent misalignment angles between the sample and the sample stage.}
\end{figure}

From Eq. (2) the expected energy shift between EDCs at the X and Y points for $\delta \simeq 1 \%$ is
\begin{eqnarray}
\varepsilon (\pi,0) - \varepsilon (0, \pi) \simeq 4t\delta = 4 \times 250 \text{ meV} \times 1 \% = 10 \text{ meV}.
\end{eqnarray}
This small shift is consistent with the fact that a clear energy shift of the band bottom was not observed in Fig. 1 of the main text within the energy resolution of our experimental setup.

\clearpage

\subsection{Effect of strain on hopping parameters}

One may suspect that the anisotropy arises only from the effect of the uniaxial strain and that the temperature dependence of the anisotropy is caused by thermal expansion. Such a possibility can be excluded by the following consideration: 
here, we consider a transfer integral between atomic orbitals of different atoms {\it i} and {\it j}. When the axis connecting both atoms is taken as the $z$-axis, the transfer integral between the atoms {\it i} and {\it j} $-t_{ij}(0,0,1) = \braket{i|h|j}$ is given by the Slater-Koster parameter $(l_il_j\mu)$, where $l_A$ is the azimuthal quantum number of atom $A$ $(A=a, b)$ and $\mu$ is the magnetic quantum number of atoms $i$ and $j$.  In the following, $l_A = 0, 1, 2,...$ and $\mu = 0, \pm1, \pm2,...$ are noted as  {\it s, p, d},... and $\sigma, \pi, \delta,...$, respectively.\par
As for the transfer integral between the $2p$ orbital of O atom and the $3d$ orbital of Cu atom, there is approximate relationship between the Slater-Koster parameter and the distance between two atoms $d$ \cite{Harrison}:
\begin{equation}
(pd \mu ) \propto d^{-3.5}. \label{eq:3.2.3}
\end{equation}
Hopping between nearest-neighbor Cu atoms can be regarded as consisting of two hopping processes from Cu to O and from O to another Cu and, therefore, the nearest-neighbor hopping parameter $t$ in Eq. (1) as a function of $d$ is given by
\begin{equation}
t { \propto (pd \mu )^2} \propto d^{-7}. \label{eq:3.2.4}
\end{equation}
If we assume that $t$ depends only on $d$ for simplicity, one can obtain
\begin{equation}
\frac{\Delta t}{t} \simeq -7\frac{\Delta d}{d}. \label{eq:3.2.5}
\end{equation}
If the anisotropy of $t$ which we observed is only due to the uniaxial strain, the anisotropy parameter $\delta (=\Delta t/t)$ should change monotonically because the atomic distance changes monotonically with temperature by thermal expansion. However, the anisotropy parameter $\delta$ observed in our measurement was enhanced only in the pseudogap state. This means that the stress works to align the nematic domain properly and ensures that the anisotropy does not come only from the distortion of the sample.
\clearpage

%